\pdfoutput=1

\documentclass[ amsmath, amssymb, aps, prd, twocolumn, lengthcheck, sort&compress, showpacs, nofootinbib ]{revtex4-1}

\usepackage{amsmath}
\usepackage{amssymb}
\usepackage{graphicx}
\usepackage{dsfont}
\usepackage{dcolumn}
\usepackage{units}
\usepackage{bm}
\usepackage{wasysym}

\usepackage[usenames]{color}

\usepackage[colorlinks=true,linkcolor=blue,citecolor=blue,urlcolor=blue]{hyperref}

      \newcommand{\conjg}[1]{\ensuremath{\hspace{1pt}\overline{\hspace{-1pt}#1\hspace{-1pt}}}\hspace{1pt}}
      \newcommand{\vect}[1]{\bm{#1}}

         \newcommand{\be}{\begin{equation}}
         \newcommand{\ee}{\end{equation}}

\def\Slash#1{\setbox0=\hbox{$#1$} 
\dimen0=\wd0 
\setbox1=\hbox{/} \dimen1=\wd1 
\ifdim\dimen0>\dimen1 
\rlap{\hbox to \dimen0{\hfil/\hfil}} 
#1 
\else 
\rlap{\hbox to \dimen1{\hfil$#1$\hfil}} 
/ 
\fi}

\def\longlongrightarrow{
\relbar\joinrel\relbar\joinrel\relbar\joinrel\relbar\joinrel\rightarrow}
\def\longlonglongrightarrow{
\relbar\joinrel\relbar\joinrel\relbar\joinrel\relbar\joinrel\relbar\joinrel\relbar\joinrel\rightarrow}
\def\longlonglonglongrightarrow{
\relbar\joinrel\relbar\joinrel\relbar\joinrel\relbar\joinrel\relbar\joinrel\relbar\joinrel\relbar\joinrel\relbar\joinrel\relbar\joinrel\rightarrow}

\begin{document}

         \title{Nucleon axial and pseudoscalar form factors \\ from the covariant Faddeev equation  }

         \author{Gernot~Eichmann}
         \author{Christian~S.~Fischer}
         \affiliation{Institut f\"{u}r Theoretische Physik, Justus-Liebig-Universit\"at Giessen, D-35392 Giessen, Germany}

         \date{\today}

         \begin{abstract}

         We compute the axial and pseudoscalar form factors of the nucleon in the Dyson-Schwinger
         approach. To this end, we solve a covariant three-body Faddeev equation for the nucleon wave function
         and determine the matrix elements of the axialvector and pseudoscalar isotriplet currents.
         Our only input is a well-established and phenomenologically successful ansatz for the nonperturbative quark-gluon interaction.
         As a consequence of the axial Ward-Takahashi identity that is respected at the quark level,
         the Goldberger-Treiman relation is reproduced for all current-quark masses.
         We discuss the timelike pole structure of the quark-antiquark vertices that enters the nucleon matrix elements and
         determines the momentum dependence of the form factors. 
         Our result for the axial charge underestimates the experimental value by $20-25\%$
         which might be a signal of missing pion-cloud contributions.
         The axial and pseudoscalar form factors agree with phenomenological and lattice data in the momentum range above $Q^2 \sim 1\dots 2$~GeV$^2$.

         \end{abstract}


         \maketitle


\section{Introduction}

              The nucleon's axial and pseudoscalar form factors are of fundamental significance for the properties
              of the nucleon that are probed in weak interaction processes. Their momentum dependence can
              be experimentally tested by (anti)neutino scattering off nucleons or nuclei, charged pion electroproduction
              and muon capture processes; see~\cite{Bernard:2001rs,Gorringe:2002xx,Schindler:2006jq} for
              reviews. Both form factors are experimentally hard to extract and therefore considerably less
              well known than their electromagnetic counterparts. Precisely measured is only the low-momentum
              limit $g_A$ of the axial form factor which is determined from neutron $\beta$-decay.
              Planned experiments at major facilities are expected to change this situation in the near future.

              The theoretical calculation of the nucleon's axial and pseudoscalar form factors requires
              genuinely non-perturbative methods. Chiral perturbation theory has been successful in this
              respect~\cite{Bernard:2001rs,Bernard:2006te,Procura:2006gq} although it is generally limited to the region of low momentum transfer.
              Recent studies in lattice gauge theory are getting closer to the physical pion mass region~\cite{Yamazaki:2009zq,Bratt:2010jn,Alexandrou:2010hf}
              but finite-volume effects become increasingly important.
              Another non-perturbative approach is the one via functional methods,
              in particular Dyson-Schwinger equations (DSEs) and Faddeev equations.
              The basic idea here is to start
              at the level of the Green functions for quarks and gluons and construct hadronic bound states from
              corresponding Bethe-Salpeter equations (BSEs) and Faddeev equations~\cite{Alkofer:2000wg,Fischer:2006ub,Chang:2011vu}.
              Those bound states are subsequently probed by the respective currents to yield the form factors of interest. One of the important advantages of
              this approach is its direct access to the substructure of mesons and baryons at all momentum
              scales and values of the current-quark mass. In principle such a setup is ideal to identify mechanisms
              that are responsible for the wealth of experimental phenomena associated with the study of hadrons.

              The study of axial and pseudoscalar form factors in the functional approach has so far been limited to
              an approximation where the nucleon is treated as a bound object of a quark and a diquark
              that interact via quark exchange~\cite{Hellstern:1997pg,Oettel:2000jj}.
              The entire gluonic substructure appears here only implicitly within the dressing of
              quark and diquark propagators as well as diquark vertex functions.
              There are several conceptual issues that complicate the treatment of form factors in the quark-diquark model.
              First, the requirement of current conservation induces the
              appearance of intricate 'seagull' diagrams~\cite{Oettel:1999gc}.
              Such terms have been taken into account for electromagnetic form factors,
              but their implementation in the case of axial form factors has not yet been possible
              for technical reasons~\cite{Oettel:2000jj}.
              Second, to comply with chiral Ward identities,
              a current-conserving quark-diquark model requires vector diquarks
              in addition to the usual scalar and axialvector diquark degrees of freedom~\cite{Ishii:2000zy}.
              Such an elaborate treatment of the quark-diquark model has not yet been performed.

              The situation is somewhat different when the nucleon is treated as a genuine
              three-body problem. The resulting Faddeev equation in rainbow-ladder truncation has been
              solved only recently for the nucleon and $\Delta$ masses~\cite{Eichmann:2009qa,SanchisAlepuz:2011jn},
              and the corresponding nucleon electromagnetic form factors have been calculated in
              Ref.~\cite{Eichmann:2011vu}. Here the currents couple to the fully dressed quark propagators
              only and thereby provide a clean access to the underlying quark and gluon substructure,
              without the additional complications of the quark-diquark model. In the present work we dwell
              on this advantage and determine the nucleon's axial and pseudoscalar form factors in the three-body approach.

              The paper is organized as follows. In Sec.~\ref{sec:Faddeev} we summarize the details of the
              three-body Faddeev framework and specify the effective quark-gluon coupling that serves as a
              model input in our calculation. We furthermore discuss the structure of the axial
              and pseudoscalar vertices that are central for our calculation of the respective form factors.
              Here we made additional progress in determining these vertices explicitly from their inhomogenuous
              Bethe-Salpeter equations instead of the model ans\"atze that were used in previous works~\cite{Oettel:2000jj}.
              In Sec.~\ref{sec:ffs} we briefly discuss the various interrelatons between
              the axial, pseudoscalar and pion-nucleon form factors and the resulting Goldberger-Treiman
              relation. Our results are presented in Sec.~\ref{sec:results} and we conclude in Sec.~\ref{sec:conclusions}.
              Throughout this paper we work in Euclidean momentum space and flavor-$SU(2)$ in the isospin-symmetric limit $m_u=m_d$.

        \begin{figure*}[t]
                    \begin{center}

                    \includegraphics[scale=0.16]{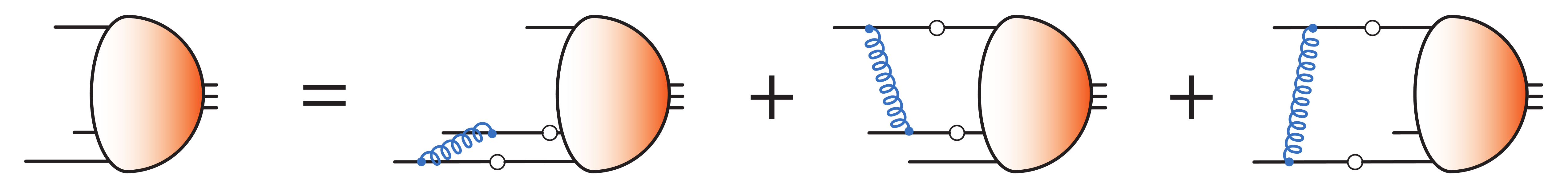}
                    \caption{(Color online) The covariant Faddeev equation for a baryon amplitude, Eq.\,(\ref{faddeev:eq}),
                    in rainbow-ladder truncation.}\label{fig:faddeev}

                    \end{center}
        \end{figure*}


\section{Faddeev framework} \label{sec:Faddeev}

        Our goal is to compute the matrix elements of the axialvector and pseudoscalar isotriplet currents 
        \begin{equation}\label{currents-general}
            \bar{q}(x)\,\gamma_5\gamma^\mu\,\frac{\sigma_k}{2}\,q(x) \quad\; \text{and}\quad\;
            \bar{q}(x)\,i\gamma_5\,\frac{\sigma_k}{2}\,q(x)
        \end{equation}
	    in the nucleon. The $\sigma_k$ represent the three Pauli matrices in $SU(2)$.
        The nucleon matrix elements obtained from Eq.~\eqref{currents-general} are expressed
        in terms of three Lorentz-invariant form factors that depend on the momentum transfer $Q^2$.
        The axial form factor $G_A(Q^2)$ and induced pseudoscalar form factor $G_P(Q^2)$ parametrize
        the axialvector current, whereas the pseudoscalar form factor $G_5(Q^2)$ determines the pseudoscalar current.
        We will discuss the respective decomposition in detail in Section~\ref{sec:ffs}.

        The computation of these form factors in the Dyson-Schwinger framework
        necessitates a quark-level picture of the nucleon current matrix elements.
        In the following subsections we will outline a description where all ingredients
        of the current diagrams are computed selfconsistently
        once the quark-(anti-)quark interaction is specified.
        The nucleon's bound-state amplitude as solution of the covariant Faddeev equation is discussed in Section~\ref{sec:faddeev} and
        the ansatz for the kernel and corresponding equation for the dressed quark propagator in Section~\ref{sec:qprop}.
        The construction of the nucleon's current matrix elements is described in Section~\ref{sec:current}, and
        in Section~\ref{sec:vertices} we detail the properties of the axialvector and pseudoscalar $q\bar{q}$
        vertices that enter the form factor diagrams.


\subsection{Covariant Faddeev equation} \label{sec:faddeev}

        The starting point of the covariant bound-state approach is the three-quark Green function $G$,
        as well as its three-quark connected and amputated counterpart, the scattering matrix $T$ that is defined via
        \begin{equation}\label{T-definition}
            G = G_0 + G_0\,T\,G_0\,.
        \end{equation}
        Here, $G_0=S\otimes S\otimes S$ denotes the disconnected contribution to $G$, i.e., the product of three dressed quark propagators.
        In order to keep the discussion transparent, we will frequently resort to a symbolic notation
        where Dirac-Lorentz, color and flavor indices as well as momentum dependencies are suppressed, and
        the products in Eq.~\eqref{T-definition} and related equations are understood as four-momentum integrations over all internal loop momenta.
        The Green function $G$ satisfies a scattering equation, which is the
        nonperturbatively resummed Dyson series
        \begin{equation}\label{scattering-eq-for-G}
             G = G_0 + G_0\,K\,G \quad \Leftrightarrow \quad G^{-1} = G_0^{-1} - K\,,
        \end{equation}
        and the equivalent relation for the $T-$matrix reads
        \begin{equation}\label{dyson-eq}
             T = K + K \,G_0 \,T \quad \Leftrightarrow \quad T^{-1} = K^{-1} - G_0\,.
        \end{equation}
        The input in both equations is the three-quark kernel $K$ that will be detailed below. 

        Baryons in QCD appear as poles in the scattering matrix $T$.
        Such a pole defines a baryon bound state on its mass shell $P^2=-M^2$,
        where $P$ is the total baryon momentum and $M$ is its mass.
        The scattering matrix at the pole assumes the form
             \begin{equation}\label{poles-in-T}
                T\stackrel{P^2=-M^2}{\longlonglongrightarrow} \frac{\Psi\,\conjg{\Psi}}{P^2+M^2}
             \end{equation}
        and thereby defines the baryon's covariant bound-state amplitude $\Psi$, with $\conjg{\Psi}$ being its charge conjugate.
             At the pole, Eq.~\eqref{dyson-eq} reduces to a homogeneous bound-state equation for the amplitude $\Psi$:
            \begin{equation}\label{three-quark-eq}
                \Psi = K\,G_0\,\Psi\,, \qquad
                K = K_\text{[3]} + \sum_{a=1}^3 S^{-1}_{(a)}\otimes K_{(a)}\,.
            \end{equation}
            The three-body kernel $K$
            contains a three-quark irreducible contribution $K_{[3]}$ and the sum of permuted two-quark kernels $K_{(a)}$,
            where the subscript $a$ denotes the respective spectator quark.

            The first step in constructing a feasible ansatz for $K$ is to disregard irreducible three-quark interactions such as,
            in the simplest case, those originating from a three-gluon vertex that connects all three quark lines.
            Omitting $K_{[3]}$ in Eq.~\eqref{three-quark-eq} yields the covariant Faddeev equation which
            traces the binding mechanism of three quarks in a baryon to its quark-quark correlations.
            It is illustrated in Fig.~\ref{fig:faddeev}, where we have already anticipated the gluon-exchange interaction
            that will be motivated in Section~\ref{sec:qprop}.
            The Faddeev equation reads explicitly:
                \begin{equation}\label{faddeev:eq}
                \begin{split}
                    \Psi_{\alpha\beta\gamma\delta}(p, q,P)  = \int  \Big[ \,
                    & \widetilde{K}_{\beta\beta'\gamma\gamma'}^{(1)} \, \Psi_{\alpha\beta'\gamma'\delta}(p^{(1)},q^{(1)},P) \,+   \\[-1mm]
                    & \widetilde{K}_{\gamma\gamma'\alpha\alpha'}^{(2)} \, \Psi_{\alpha'\beta\gamma'\delta}(p^{(2)},q^{(2)},P) \, + \\
                    & \widetilde{K}_{\alpha\alpha'\beta\beta'}^{(3)} \, \Psi_{\alpha'\beta'\gamma\delta}(p^{(3)},q^{(3)},P) \,\Big] \,.
                \end{split}
                \end{equation}
            Here, $\widetilde{K}^{(a)}$ abbreviates the renormalization-group invariant product of two dressed quark propagators
            and the remaining two-quark kernel $\mathcal{K}$:
            \begin{equation} \label{KSS}
                \widetilde{K}_{\alpha\alpha'\beta\beta'}^{(a)} = \mathcal{K}_{\alpha\alpha''\beta\beta''}\,
                                                                  S^{(b)}_{\alpha''\alpha'} \, S^{(c)}_{\beta''\beta'} \,,
            \end{equation}
            where $\{a,b,c\}$ is an even permutation of $\{1,2,3\}$.
            The covariant nucleon amplitude $\Psi$ carries three spinor indices $\{\alpha,\beta,\gamma\}$ for the valence quarks
            and one index $\delta$ for the spin-$1/2$ nucleon.
            It depends on two relative momenta $p$ and $q$
            and the total nucleon momentum $P$, where $P^2 = -M^2$ is fixed.
            For details on the kinematics and notation, as well as the solution method, we refer the reader to
            Refs.~\cite{Eichmann:2009qa,Eichmann:2009en,Eichmann:2011vu}.

            It is noteworthy that Poincar\'e covariance allows for a rich
            spin and quark orbital angular-momentum structure in the three-quark Faddeev amplitude.
            Specifically, the nucleon's bound-state amplitude
            consists of 64 covariant basis elements which can be classified in sets of $s-$, $p-$ and $d-$waves
            in the nucleon's rest frame~\cite{Eichmann:2009qa}.
            While the nucleon ground state is dominated by $s-$waves, a gluon-exchange
            interaction generates $\sim 30\%$ $p-$wave admixture in the nucleon's canonical normalization, and $d-$waves
            contribute  roughly $1\%$~\cite{Eichmann:2011vu}.

\subsection{Quark propagator and $qq$ kernel} \label{sec:qprop}

            The kernel~\eqref{KSS} of the covariant Faddeev equation includes the two crucial ingredients
            that relate the properties of hadrons to the underlying structure of QCD:
            the dressed quark propagator $S(p)$ and the two-quark kernel $\mathcal{K}$ which is irreducible with respect to a $qq$ pair.
            Repeating the steps~(\ref{dyson-eq}--\ref{three-quark-eq}) in the quark-antiquark channel yields
            the homogeneous Bethe-Salpeter equation (BSE) for a meson that
            involves the quark-antiquark analogue of the kernel $\mathcal{K}$.
            The same interaction thereby enters both $q\bar{q}$ and $qqq$ bound-state equations which emphasizes
            the close connection between meson and baryon properties in our approach.

            To ensure the pion's nature as the Goldstone boson of spontaneous chiral symmetry breaking,
            the axial-vector Ward-Takahashi identity (AXWTI) must be satisfied. The AXWTI
            entails a massless pion in the chiral limit and leads to a generalized Gell-Mann--Oakes--Renner
            relation~\cite{Maris:1997hd,Holl:2004fr}, and it poses a constraint on the construction of the
            kernel $\mathcal{K}$ by relating it with the kernel of the quark DSE, cf.~Eq.~\eqref{axwti-kernel}.
            The simplest kernel to satisfy that constraint is
            the rainbow-ladder truncation which amounts to
                \begin{equation}\label{RLkernel}
                    \mathcal{K}_{\alpha\alpha'\beta\beta'} =  Z_2^2 \, \frac{ 4\pi \alpha(k^2)}{k^2} \,
                    T^{\mu\nu}_k \gamma^\mu_{\alpha\alpha'} \,\gamma^\nu_{\beta\beta'},
                \end{equation}
            where $T^{\mu\nu}_k=\delta^{\mu\nu} - k^\mu k^\nu/k^2$
           is a transverse projector with respect to the gluon momentum $k$ and $Z_{2}$ is the quark renormalization constant.
           Eq.~\eqref{RLkernel} describes an iterated dressed-gluon exchange between quark and antiquark that retains
           only the vector part $\sim \gamma^\mu$ of the quark-gluon vertex.
           Its non-perturbative dressing, together with that of the gluon propagator,
           is absorbed into an effective coupling $\alpha(k^2)$ which is modeled.

           There have been various attempts to go beyond the rainbow-ladder truncation by employing different strategies. While
           Refs.~\cite{Watson:2004kd,Alkofer:2008tt,Alkofer:2008et,Fischer:2009jm,Fischer:2008wy} analyzed infrared and pion-cloud contributions
           in the quark-gluon interaction directly from the vertex DSE,
           Refs.~\cite{Chang:2011vu,Chang:2009zb,Chang:2010hb} use a Ward-Takahashi identity for the Bethe-Salpeter kernel as a
           guidance for vertex models. These approaches have been quite successful in the light meson sector but their implementation
           for baryons is still hampered by technical difficulties. For this reason we restrict ourselves to the well-established
           rainbow-ladder kernel of Eq.~\eqref{RLkernel}. Its main drawback appears in the form of missing
           pion-cloud contributions which are important for the form factors in the chiral and low-momentum region~\cite{Eichmann:2008ae,Eichmann:2011vu}.
           This issue is discussed in more detail in section \ref{sec:results}.

           The second ingredient in the Faddeev kernel is the dressed quark propagator.
           It is expressed in terms of two scalar functions, the quark wave-function renormalization $1/A(p^2)$ and the quark mass function $M(p^2)$:
                  \begin{equation}\label{qprop}
                      S^{-1}(p) = A(p^2)\,\left( i\Slash{p} + M(p^2) \right).
                  \end{equation}
           The quark propagator satisfies the quark DSE whose interaction kernel includes the dressed gluon propagator as well as one bare
      	  and one dressed quark-gluon vertex. In rainbow-ladder truncation that kernel becomes identical to Eq.~\eqref{RLkernel} and the quark DSE reads
                  \begin{equation}\label{quarkdse}
                      S^{-1}_{\alpha\beta}(p) = Z_2 \left( i\Slash{p} + m_0 \right)_{\alpha\beta}  +
      		               \int \mathcal{K}_{\alpha\alpha'\beta'\beta} \,S_{\alpha'\beta'}(q)\,.
                  \end{equation}
          The bare current-quark mass $m_0$ is related to the renormalized current mass $m_q$ via
          \begin{equation}
              Z_2  m_0 = Z_2 Z_m m_q = Z_4  m_q\,,
          \end{equation}
          where the quark-mass and wave-function renormalization constants $Z_m$ and $Z_2$
          are determined self-consistently in the process of solving the quark DSE.
          The current-quark mass
          constitutes an input of the equation and can be readily varied from the chiral limit up to the heavy-quark regime.

          Dynamical chiral symmetry breaking becomes manifest in the quark DSE if the kernel $\mathcal{K}$ supplies sufficient interaction strength.
          The consequence is a non-perturbative enhancement of the quark mass function $M(p^2)$ at small momenta that
          indicates the dynamical generation of a constituent-quark mass scale.
          In principle, such strength would be generated through a self-consistent DSE solution for the gluon propagator and quark-gluon vertex
          that enter Eq.~\eqref{quarkdse}; see, e.g.,~\cite{Fischer:2006ub,Chang:2011vu} and references therein.
          In rainbow-ladder that effect is provided by the effective coupling $\alpha(k^2)$ which we choose
          from Ref.~\cite{Maris:1999nt} as\footnote{We do not expect much difference in our results from a recently
           suggested update~\cite{Qin:2011dd} of Eq.~\eqref{couplingMT}.}
                        \begin{equation}\label{couplingMT}
                            \alpha(k^2) = \pi \eta^7  \left[\frac{k^2}{\Lambda^2}\right]^2 \!\!
                            e^{-\eta^2 \left[\frac{k^2}{\Lambda^2}\right]} + \alpha_\text{UV}(k^2) \,.
                        \end{equation}
            The second term is only relevant at large gluon momenta where it dominates and is constrained by perturbative QCD, i.e.,
            it preserves the one-loop renormalization group behavior of QCD for solutions of the quark DSE:
            \begin{equation}
                \alpha_\text{UV}(k^2) = \frac{2\pi\gamma_m \big(1-e^{-k^2/\Lambda_t^2}\big)}{\ln\,[e^2-1+(1+k^2/\Lambda^2_\mathrm{QCD})^2]}\,,
            \end{equation}
            with $\Lambda_t=1$~GeV, $\Lambda_\mathrm{QCD}=0.234\,{\rm GeV}$, and $\gamma_m=12/25$.
            The first term provides the necessary strength at small and intermediate momenta
            that triggers the transition from a current-quark to a dynamically generated constituent quark.
            It is characterized by two parameters\footnote{The
            relationship with the parameters $\{ \omega, D \}$ used in Ref~\cite{Maris:1999nt} is
            $\omega D=\Lambda^3$, $\eta = \Lambda/\omega$.}:
            an infrared scale $\Lambda$ that represents the scale of dynamical chiral symmetry breaking,
            and a dimensionless width parameter $\eta$.

        In combination with the interaction of Eq.~\eqref{couplingMT}, the rainbow-ladder truncation
        has been frequently used in Dyson-Schwinger studies of hadrons.
        By setting the scale $\Lambda$ via the experimental pion decay constant,
        the approach describes pseudoscalar-meson, vector-meson,
        nucleon and $\Delta$ ground-state observables reasonably well, see
        e.g.~\cite{Maris:2005tt,Maris:2006ea,Krassnigg:2009zh,Nicmorus:2010mc} and references therein.
        In addition, their properties are insensitive to a variation of the infrared shape of the
        coupling~\cite{Maris:1999nt,Krassnigg:2009zh} which is controlled by the parameter $\eta$.

    \subsection{Nucleon current} \label{sec:current}

        After having specified the basic ingredients that enter the description of baryons in the bound-state approach, we turn to
        the question of resolving the nucleon matrix elements of the currents in Eq.~\eqref{currents-general} in terms of QCD's non-perturbative Green functions.
        A systematic construction of the nucleon's axial and pseudoscalar current in the three-quark framework
        is provided by the 'gauging of equations' method of Refs.~\cite{Haberzettl:1997jg,Kvinikhidze:1998xn,Kvinikhidze:1999xp}.
        It was previously applied to derive the electromagnetic current in the quark-diquark model~\cite{Oettel:1999gc,Oettel:2000ig} as well as
        in the present three-quark setup~\cite{Eichmann:2011vu},
        and a recent computation of the pseudoscalar $N-\Delta$ transition current is based on that procedure as well~\cite{Mader:2011zf}.
        Moreover, the approach can be generalized to yield a quark-level description of generalized parton distributions~\cite{Kvinikhidze:2004dy} and
        a variety of hadronic scattering processes~\cite{Eichmann:2011ec}.
        Since the method is not partial to the type of the involved current, we can simply adopt the derivation in Ref.~\cite{Eichmann:2011vu}
        to the case investigated here.

        The coupling of an external current to a baryon amounts to 'gauging' the $qqq$ scattering matrix $T$.
        That operation acts as a derivative, i.e., it is linear and satisfies Leibniz' rule, and we formally denote it by a superscript $[\mu]$.
        The current $J^{[\mu]}$ is the residue of the gauged scattering matrix $T^{[\mu]}$ at the hadron's bound-state pole:
             \begin{equation}
                 T^{[\mu]} \stackrel{P_i^2=P_f^2=-M^2}{\longlonglonglongrightarrow}   - \frac{\Psi_f\,J^{[\mu]} \,\conjg{\Psi}_i}{(P_f^2+M^2)(P_i^2+M^2)}\,,
             \end{equation}
             where $\Psi_i = \Psi(p_i,q_i,P_i)$ and $\Psi_f = \Psi(p_f,q_f,P_f)$ are in- and outgoing baryon amplitudes with different
             momentum dependencies.
             From the derivative property $T^{[\mu]} = - T \left(T^{-1}\right)^{[\mu]} T$,
             in combination with Eqs.~(\ref{dyson-eq}--\ref{poles-in-T}),
             one derives the general expression for a baryon's nonperturbative current.
             It is the gauged inverse T-matrix sandwiched between the bound-state amplitudes:
             \begin{equation}\label{emcurrent-gauging}
             \begin{split}
                J^{[\mu]} &= \conjg{\Psi}_f \left(T^{-1}\right)^{[\mu]} \Psi_i = \\
                          &= \conjg{\Psi}_f \,G_0 \left(\mathbf{\Gamma}^{[\mu]} - K^{[\mu]} \right) G_0 \,\Psi_i \,.
             \end{split}
             \end{equation}
             In the second step we have exploited Eq.~\eqref{dyson-eq},
             \begin{equation}\label{T-inv-gauged}
             \begin{split}
                \left(T^{-1}\right)^{[\mu]} &= \left( K^{-1}-G_0\right)^{[\mu]} = \\
                                        &= G_0 \left(G_0^{-1}\right)^{[\mu]} G_0 - K^{-1}\,K^{[\mu]}\,K^{-1} \,,
             \end{split}
             \end{equation}
             and defined the 'three-body' vertex
             \begin{equation}
                 \mathbf{\Gamma}^{[\mu]}:=\left(G_0^{-1}\right)^{[\mu]}= \left(S^{-1}\otimes S^{-1}\otimes S^{-1}\right)^{[\mu]}
             \end{equation}
             as the gauged disconnected inverse propagator product.
             To arrive at Eq.~\eqref{emcurrent-gauging},
             we have eliminated the inverse kernels that appear in Eq.~\eqref{T-inv-gauged} by inserting the bound-state equations
             $K^{-1} \Psi_i = G_0 \Psi_i$ and $\conjg{\Psi}_f K^{-1} = \conjg{\Psi}_f \,G_0$.
             The resulting expression for the current matrix element involves the same elements that appear in the bound-state equation~\eqref{three-quark-eq},
             namely the dressed quark propagator and the three-quark kernel $K$.

\begin{figure}[t]
            \begin{center}
            \includegraphics[scale=0.11]{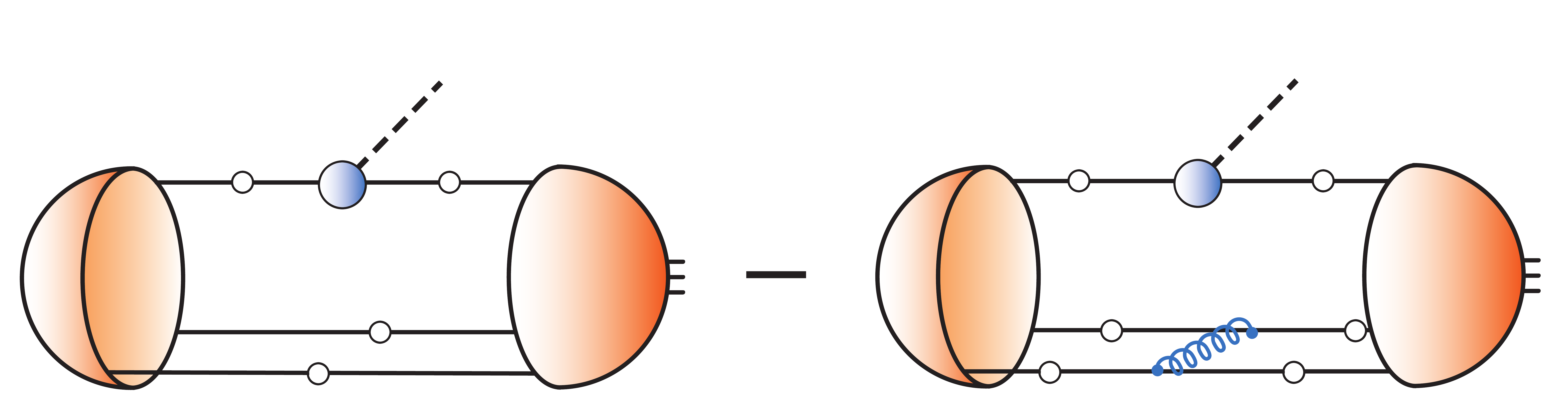}
            \caption{(Color online)
                     The two types of diagrams (modulo permutations) which contribute to the nucleon's three-body current
                     in rainbow-ladder truncation, Eqs.~\eqref{emcurrent-gauging} and~\eqref{current-ingredients}.} \label{fig:current}
            \end{center}
            \end{figure}

             The question remains what 'gauging' of $S^{-1}$ and $K$ means at the quark level. Microscopically,
             the axialvector and pseudoscalar couplings to the quark are represented
             by the respective quark-antiquark vertices $\Gamma^{[\mu]}$.
             We can identify them with the gauged inverse quark propagator, i.e., $\left(S^{-1}\right)^{[\mu]} = \Gamma^{[\mu]}$,
             and correspondingly: $S^{[\mu]} =-S \,\Gamma^{[\mu]} \,S$.
             The gauged three-body kernel $K^{[\mu]}$ is obtained by applying the Leibniz rule to Eq.~\eqref{three-quark-eq}.
             In rainbow-ladder truncation, where the three-body irreducible contribution to $K$ is neglected
             and the two-body kernel is modeled by gluon exchange, only the direct couplings to the quarks remain.
            Then, the ingredients of Eq.~\eqref{emcurrent-gauging} become
            \begin{equation}\label{current-ingredients}
            \begin{split}
                   \mathbf{\Gamma}^{[\mu]} &= \sum_{a=1}^3 \Gamma^{[\mu]}_{(a)} \otimes S^{-1}_{(b)} \otimes S^{-1}_{(c)}\,,\\
                   K^{[\mu]} &= \sum_{a=1}^3 \Gamma^{[\mu]}_{(a)} \otimes K_{(a)} \,,
            \end{split}
            \end{equation}
            where the quark labels $\{a,b,c\}$ are again an even permutation of $\{1,2,3\}$.
            The resulting current is illustrated in Fig.~\ref{fig:current}
            and consists of an impulse-approximation diagram and another contribution that involves the gluon-exchange kernel.
            These diagrams are worked out in detail in App.~\ref{sec:emcurrent-worked-out}.

            Here we find another practical advantage of the rainbow-ladder truncation:
            due to the structure of Eqs.~\eqref{emcurrent-gauging} and~\eqref{current-ingredients},
            all ingredients of the nucleon's current are already specified, namely by the ansatz~\eqref{RLkernel} alone.
            This is also true for the dressed axialvector and pseudoscalar vertices that appear in Fig.~\ref{fig:current},
            and we will discuss them in the following subsection.

    \subsection{Axial and pseudoscalar vertices} \label{sec:vertices}

            In order to compute the form factor diagrams of Fig.~\ref{fig:current} we need to specify the
            dressed pseudoscalar and axialvector quark-antiquark vertices, i.e., the
            'gauged' quark propagators, that enter the diagrams.
            Their properties have been established and discussed in detail in Refs.~\cite{Maris:1997hd,Maris:1999bh,Holl:2005vu,Blank:2011qk,Chang:2011vu}.
            Since they have a direct correspondence with the nucleon's axial and pseudoscalar form factors,
            we find it useful to recollect some of these features here and also draw attention to a few properties that
            are perhaps less well-known.

\begin{figure*}[t]
                    \begin{center}

                    \includegraphics[scale=0.118]{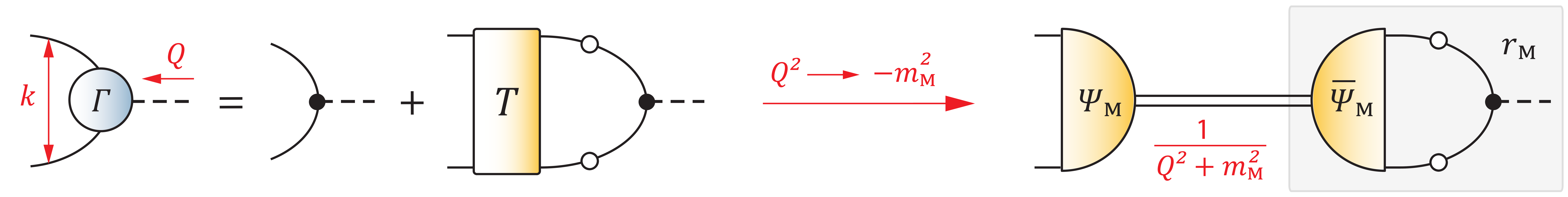}
                    \caption{(Color online) Diagrammatical representation of Eq.~\eqref{G-def} and its onshell limit,
                    Eqs.~(\ref{G-pole}--\ref{r-def}).}\label{fig1}

                    \end{center}
        \end{figure*}

            We can express the coupling of a dressed quark to a current with pseudoscalar, vector or axialvector tensor structure $\Gamma_0$, with
            \begin{equation}
                \Gamma_0 \in \left\{ \, Z_4 i\gamma_5\,, \;  Z_2  i\gamma^\mu\,, \;  Z_2 \gamma_5 \gamma^\mu \,\right\},
            \end{equation}
            by the quark-antiquark vertex $\Gamma^{[\mu]}$, where the gauging index characterizes the
            type of $\Gamma_0$. The vertex depends on two momenta,
            the relative $q\bar{q}$ momentum $k$ and the total momentum $Q$ (see Fig.~\ref{fig1}).
            We consider isotriplet currents with flavor matrices $\sigma_k/2$, cf.~Eq.~\eqref{currents-general}.
            The vertex $\Gamma^{[\mu]}$ can be defined as the contraction of the
            quark-antiquark four-point function $G$ with $\Gamma_0$:
            \begin{equation}\label{G-def}
                \Gamma^{[\mu]} = G_0^{-1}\,G\,\Gamma_0 = \Gamma_0 + T\,G_0\,\Gamma_0\,,
            \end{equation}
            where we have exploited Eq.~\eqref{T-definition} in the quark-antiquark case,
            and $G_0=S\otimes S$ is here the product of two dressed quark propagators.
            This relation is illustrated in Fig.~\ref{fig1}.
            The scattering equation~\eqref{dyson-eq} immediately yields an inhomogeneous BSE for $\Gamma^{[\mu]}$,
            \begin{equation}\label{ibse}
                \Gamma^{[\mu]} = \Gamma_0 + K\,G_0\,\Gamma^{[\mu]}\,,
            \end{equation}
            which allows to compute the vertex for a given kernel $K$.

            The quark-antiquark scattering matrix $T$ incorporates all meson poles,
            and on the mass shell $Q^2 = -m_\text{M}^2$ for a meson with mass $m_\text{M}$ it assumes the form
            \begin{equation}\label{G-pole}
                T \stackrel{Q^2 \rightarrow -m_\text{M}^2}{\longlonglongrightarrow} \frac{\Psi_\text{M}
                \,\conjg{\Psi}_\text{M}}{Q^2+m_\text{M}^2}     \,,
            \end{equation}
            where $\Psi_\text{M}$ defines the meson's homogeneous bound-state amplitude.
            The central feature that will have an impact on the momentum structure of hadron form factors is the fact that
            these poles, via Eq.~\eqref{G-def}, also appear in the vertex $\Gamma^{[\mu]}$, cf.~Fig.~\ref{fig1}:
            \begin{equation}\label{G-Pole2}
                \Gamma^{[\mu]} \stackrel{Q^2 \rightarrow -m_\text{M}^2}{\longlonglongrightarrow} \frac{r_\text{M}}{Q^2+m_\text{M}^2}\,\Psi_\text{M}\,.
            \end{equation}
            This is true
            as long as the bound-state wave function $G_0 \Psi_\text{M}$ has a component in the direction of $\Gamma_0$,
            which is expressed by the residue
            \begin{equation}\label{r-def}
                r_\text{M}[\Gamma_0] = \text{Tr}\int_k\conjg{\Psi}_\text{M}\,G_0\,\Gamma_0 \,\Big|_{Q^2\rightarrow -m_\text{M}^2}\,.
            \end{equation}

            For example, the pseudoscalar vertex $\Gamma_5$ contains all pseudoscalar meson poles, and the
            quark-photon (i.e., vector) vertex $\Gamma^\mu$ inherits the vector-meson poles. The
            respective residues for the pion and the $\rho-$meson are given by~\cite{Maris:1997hd,Maris:1999bh}
            \begin{align}
                \Gamma_0 = Z_4 i\gamma_5 & \quad\Rightarrow\quad     r_\pi[\Gamma_0]= \frac{f_\pi m_\pi^2}{2m_q}\,,  \label{rpi}  \\
                \Gamma_0 = Z_2 i\gamma^\mu & \quad\Rightarrow\quad   r^{\mu\nu}_\rho[\Gamma_0]= -f_\rho m_\rho\,T_Q^{\mu\nu} \,, \label{rrho}
            \end{align}
            where $f_\pi$ and $f_\rho$ are the pion and $\rho-$meson leptonic decay constants and $m_q$ is the renormalized
            current-quark mass.

            Turning to the axialvector vertex $\Gamma_5^\mu$, it
            is advantageous to write the vertex as a sum of purely transverse and longitudinal contributions with respect to the total momentum $Q$:
                  \begin{equation}\label{vertex:AX-TL}
                      \Gamma^\mu_5(k,Q) =   \Gamma^\mu_{5,T}(k,Q) + \frac{Q^\mu}{Q^2}\,\Gamma_{5,L}(k,Q)\,,
                  \end{equation}
            so that $Q^\mu \,\Gamma^\mu_5 = \Gamma_{5,L}$.
            Analyticity at $Q^2 = 0$ implies correlations between the transverse and longitudinal parts, see Eq.~\eqref{analyticity} below.
            Since axialvector mesons are transverse, their bound-state poles will appear only in $\Gamma^\mu_{5,T}$ whereas the longitudinal part
            $\Gamma_{5,L}$ contains pseudoscalar poles. This can be seen from inserting
            \begin{equation}\label{gammam-transverse}
                \gamma_5\gamma^\mu = \gamma_5\gamma_T^\mu +  \frac{Q^\mu}{Q^2}\,\gamma_5\Slash{Q}
            \end{equation}
            together with the decomposition~\eqref{vertex:AX-TL} into Eq.~\eqref{G-def}, upon which the equation decouples:
            \begin{equation}
            \begin{split}
                \Gamma^\mu_{5,T} &= (1+T\,G_0)\,(Z_2  \gamma_5 \gamma^\mu_T)\,, \\
                \Gamma_{5,L} &= (1+T\,G_0)\,(Z_2  \gamma_5 \Slash{Q} )\,.
            \end{split}
            \end{equation}
            From Eq.~\eqref{r-def} one infers that the longitudinal tensor component $\gamma_5 \Slash{Q}$
            can overlap with pseudoscalar pole structures encoded in $T$, for example
            the pion's bound-state wave function, and that overlap is just the definition of the pion decay constant~\cite{Maris:1997hd}:
            \begin{equation}\label{fpi-def}
                \Gamma_0 = Z_2 \gamma_5\Slash{Q}   \quad\Rightarrow\quad     r_\pi[\Gamma_0]= -f_\pi m_\pi^2\,.
            \end{equation}
            Thus, $\Gamma_{5,L}$ will exhibit pseudoscalar poles at timelike values of the total momentum-squared $Q^2$,
            and axialvector poles will appear in $\Gamma^\mu_{5,T}$.
            Similarly, the inhomogeneous axialvector BSEs~\eqref{ibse} for those two vertices decouple as well, and the longitudinal
            equation is identical to the pseudoscalar BSE except for the different driving term $Z_2 \gamma_5 \Slash{Q}$ instead of $Z_4 i\gamma_5$.
            The pole behavior will therefore be recovered in the solution of the inhomogeneous BSEs,
            and by implementing these vertices in the axial current of the nucleon, it will
            translate to the nucleon's axial form factors in the timelike region, cf.~Section~\ref{sec:ffs}.

            The precise relation between the pseudoscalar vertex and the longitudinal part of the axialvector vertex is expressed by the AXWTI
            which reads in the flavor-triplet case:
            \begin{equation}\label{ax-wti}
            \begin{split}
                      &\Gamma_{5,L}(k,Q) + 2m_q\,\Gamma_5(k,Q) = \\
                      & \quad = S^{-1}(k_+) \,i\gamma_5 + i\gamma_5 \,S^{-1}(k_-) \,,
            \end{split}
            \end{equation}
            where $k_\pm = k \pm Q/2$ are the two quark momenta.
            Eq.~\eqref{ax-wti} entails that the pseudoscalar poles
            on the left-hand side must compensate each other,
            and that the longitudinal part of the axialvector vertex is completely
            specified by the pseudoscalar vertex and the dressed quark propagator alone.

            The AXWTI can be made explicit in a given basis decomposition of the vertices.
             The transverse part of the axialvector vertex involves eight tensor structures which can be chosen as~\cite{Ball:1980ay}:
             \begin{equation}\label{transverse-vertex}
             \begin{split}
                     \Gamma^\mu_{5,T} = \gamma_5 \Big[ &  \,\gamma^\mu_T\,\big( f_1  + if_2\, k\!\cdot\! Q\,\Slash{Q} \big) + \\
                                  & + if_3\, \textstyle\frac{1}{2}  \left[\gamma^\mu_T, \,\Slash{k} \right]
                                    +  f_4\,\textstyle\frac{1}{2}  \left[\gamma^\mu_T, \,\Slash{k}_T \right] \Slash{Q} \,  + \\
                                  & + k^\mu_T \,k\!\cdot\! Q\,\big( i f_5 + f_6\,\Slash{Q} \big) + \\
                                  & + k^\mu_T \,\big( f_7\,\Slash{k}  + i f_8 \,k\!\cdot\! Q\,\Slash{k}_T \, \Slash{Q} \big) \,\Big]\,,
             \end{split}
             \end{equation}
             where the $f_i(k^2, \,k\cdot Q, \,Q^2)$ are scalar dressing functions and
             $\gamma^\mu_T$, $k^\mu_T$ are transverse with respect to the photon momentum $Q$.
             The angular prefactors $k \cdot Q$ were attached to guarantee positive charge-conjugation parity (for quarks with equal mass),
             so that the vertex carries the quantum numbers $J^{PC}=1^{++}$.
             The factors $i$ ensure that all dressing functions are real if $k^2 \in \mathds{R}_+$ and $Q^2 \in \mathds{R}$.
             Likewise, the longitudinal part can be decomposed as
             \begin{equation}\label{longitudinal-vertex}
                     \Gamma_{5,L} = i\gamma_5 \Big[ \,g_1  + i g_2\, \Slash{Q}  + i g_3\,k\!\cdot\! Q\,\Slash{k} + g_4\,\Slash{k}_T\,\Slash{Q} \Big]\,,
             \end{equation}
             and an analogous decomposition holds for the pseudoscalar $0^{-+}$ vertex:
             \begin{equation}\label{pseudoscalar-vertex}
                     \Gamma_5 = i\gamma_5 \Big[ \,h_1  + i h_2\, \Slash{Q}  + i h_3\,k\!\cdot\! Q\,\Slash{k} + h_4\,\Slash{k}_T\,\Slash{Q} \Big]\,.
             \end{equation}
             The AXWTI \eqref{ax-wti} yields the following relations between the components of $\Gamma_{5,L}$ and $\Gamma_5$
             which are valid for all current-quark masses:
             \begin{equation}\label{axwti-dressingfunctions}
             \begin{split}
                 g_1 + 2m_q\,h_1 &= 2 \,\Sigma_B\,, \\
                 g_2 + 2m_q\,h_2 &= -\Sigma_A \,, \\
                 g_3 + 2m_q\,h_3 &= -2 \,\Delta_A\,, \\
                 g_4 + 2m_q\,h_4 &= 0\,.
             \end{split}
             \end{equation}
             Here,
             $A(k^2)$ and $B(k^2)=M(k^2)A(k^2)$ are the dressing functions of the inverse quark propagator
             $S^{-1}(k)=i \Slash{k}\,A(k^2) + B(k^2)$, and we used the abbreviations
                 \begin{equation*}\label{QPV:sigma,delta}
                     \Sigma_F := \frac{F(k_+^2)+F(k_-^2)}{2} , \quad  \Delta_F := \frac{F(k_+^2)-F(k_-^2)}{k_+^2-k_-^2}\,,
                  \end{equation*}
             with $F \in \{A,B\}$.

             In the limit $Q^2\rightarrow 0$, regularity leads to an additional relation between the longitudinal and the transverse dressing functions in
             Eqs.~(\ref{transverse-vertex}--\ref{longitudinal-vertex}):
             \begin{equation}\label{analyticity}
                 g_1 = g_2+f_1 = g_3+f_7 = g_4+f_3 = 0\,,
             \end{equation}
             and thereby correlates four of the eight dressing functions of the axial-vector vertex with those of the pseudoscalar vertex
             via Eq.~\eqref{axwti-dressingfunctions}. For instance, one obtains at $Q^2\rightarrow 0$: $h_1(k^2) = B(k^2)/m_q$.
             With Eqs.~\eqref{G-Pole2} and~\eqref{rpi}, the chiral-limit behavior of the pseudoscalar vertex then becomes
             \begin{equation}
                 \Gamma_5 \stackrel{Q^2 \rightarrow  0}{\longlongrightarrow} i\gamma_5 \bigg[\, \frac{B(k^2)}{m_q} + \dots \bigg]  \stackrel{m_\pi^2 \rightarrow  0}{\longlongrightarrow} \frac{f_\pi}{2m_q}\,\Psi_\pi\,,
             \end{equation}
             where $\Psi_\pi$ is the pion's canonically normalized bound-state amplitude, with the same structure as Eq.~\eqref{pseudoscalar-vertex} except for
             the flavor convention $\sigma_k$ instead of $\sigma_k/2$. This yields the well-known chiral-limit relation
             between the dominant pion dressing function and the scalar dressing of the quark propagator~\cite{Maris:1997hd}:
             \begin{equation}\label{hpi-chiral-limit}
                 h_1^{(\pi)}(k^2)=\frac{B(k^2)}{f_\pi} \,.
             \end{equation}

            These considerations finally provide a quick way to check the Gell-Mann-Oakes-Renner relation which is already implicit in Eq.~\eqref{rpi}.
            If Eq.~\eqref{fpi-def} defines the pion decay constant,
            equating the residues of the vertices $\Gamma_{5,L}$ and $\Gamma_5$ in the AXWTI~\eqref{ax-wti} yields the quoted value for $r_\pi$ in Eq.~\eqref{rpi}.
            On the other hand, the trace with $Z_4 i\gamma_5$ in~\eqref{rpi} filters out the dominant component of the pion's wave function $G_0 \Psi_\pi$ which,
            in the chiral limit, satisfies a relation analogous to Eq.~\eqref{hpi-chiral-limit}, except that the inverse dressing $B(k^2)$
            is replaced by the scalar dressing of the quark propagator itself. Its integrated version is just
            the quark condensate, i.e., the trace over the chiral-limit quark propagator.

            While the properties discussed so far are exact in QCD,
            a practical calculation of the vertices $\Gamma^\mu_5$ and $\Gamma_5$ from the
            inhomogeneous BSE~\eqref{ibse} requires an ansatz for the kernel $K$ to operate with.
            Since the characteristics of the pion as the Goldstone boson of spontaneous chiral symmetry breaking are intimately related with the AXWTI,
            it is vital that any kernel ansatz complies with that relation.
            Through the inhomogeneous BSE~\eqref{ibse}, the AXWTI can be reformulated as a relation
            between the dressed quark propagator and the $q\bar{q}$ kernel $K$:
            \begin{equation}\label{axwti-kernel}
            \begin{split}
                & \int_{k'}\mathcal{K}_{\alpha\alpha'\beta'\beta}\left[S(k'_+)\,i\gamma_5 + i\gamma_5\,S(k'_-)\right]_{\alpha'\beta'}  \\
                & \qquad\quad = -\left[\Sigma(k_+)\,i\gamma_5 + i\gamma_5\,\Sigma(k_-)\right]_{\alpha\beta}\,,
            \end{split}
            \end{equation}
            where $\Sigma(k)$ is the quark self-energy (not to be confused with the function $\Sigma_F$ from above)
            that appears in the quark DSE~\eqref{quarkdse}: $S^{-1}(k) = Z_2\,(i\Slash{k}+m_0) + \Sigma(k)$.
            Eq.~\eqref{axwti-kernel} and analogues thereof can be used as a construction principle for suitable kernel ans\"atze
            \cite{Munczek:1994zz,Bender:1996bb},
            and it is not difficult to prove that the relation is preserved by the rainbow-ladder kernel of Eq.~\eqref{RLkernel}.

             We further recall that similar considerations apply for the quark-photon vertex $\Gamma^\mu$ whose longitudinal part,
             defined in analogy to Eq.~\eqref{vertex:AX-TL},
             must reproduce the electromagnetic Ward-Takahashi identity:
             \begin{equation}\label{em-wti}
                   Q^\mu \Gamma^\mu(k,Q) = \Gamma_L(k,Q) = S^{-1}(k_+)-S^{-1}(k_-)\,.
             \end{equation}
             Thus, the longitudinal part of the vector vertex is completely fixed by electromagnetic gauge invariance,
             i.e. by the dressed quark propagator alone. Combined with the analyticity requirement at $Q^2=0$, Eq.~\eqref{em-wti} leads to the
             Ball-Chiu construction for the vertex~\cite{Ball:1980ay}, augmented by a purely transverse term
             that includes an array of vector-meson poles. Upon implementation in the form factor diagrams,
             the latter is responsible for the timelike pole structure in electromagnetic form factors and the underlying reason for vector-meson dominance.
             Similarly to Eq.~\eqref{axwti-kernel}, the electromagnetic WTI
             can be formulated as a relation between the $q\bar{q}$ kernel and the quark self-energy:
            \begin{equation}
            \begin{split}
                & \int_{k'}K_{\alpha\alpha'\beta'\beta}\left[S(k'_+)-S(k'_-)\right]_{\alpha'\beta'}  \\
                & \qquad\qquad = \left[\Sigma(k_+)-\Sigma(k_-)\right]_{\alpha\beta}\,,
            \end{split}
            \end{equation}
            which is also satisfied by the rainbow-ladder kernel and consequently by the vector vertex as a solution of its inhomogeneous BSE.

            We have now collected all ingredients for solving
            the inhomogeneous Bethe-Salpeter equations
            for the axialvector and pseudoscalar vertices. They take the explicit form
             \begin{equation}\label{inhom-bse-detail}
             \begin{split}
                 \Gamma_{5,T}^\mu(k,Q) &= Z_2\,\gamma_5 \gamma_T^\mu - \mathcal{I} \,\big[\Gamma_{5,T}^\mu\big]\, ,\\
                 \Gamma_{5,L}(k,Q) &= Z_2\,\gamma_5\,\Slash{Q} - \mathcal{I}\,\big[\Gamma_{5,L}\big]\, , \\
                 \Gamma_5(k,Q) &= Z_4\,i\gamma_5 - \mathcal{I}\,\big[\Gamma_5\big]\, ,
             \end{split}
             \end{equation}
             where the integrals on the right-hand side are given by
             \begin{equation}\label{inhom-bse-detail-2}
                 \mathcal{I}_{\alpha\beta}\left[\Gamma\right] = \int\limits_{k'} \mathcal{K}_{\alpha\alpha'\beta'\beta}\,
                   \left[ S(k_+') \,\Gamma(k',Q)\,S(k_-')\right]_{\alpha'\beta'}\,,
             \end{equation}
             with the rainbow-ladder kernel $\mathcal{K}$ from Eq.~\eqref{RLkernel}.
             Since the rainbow-ladder truncation obeys the AXWTI per construction,
             the relations~(\ref{axwti-dressingfunctions}--\ref{analyticity}) for the resulting vertex dressing functions
             are numerically satisfied to high accuracy.
             With the axialvector and pseudoscalar vertices obtained from Eqs.~\eqref{inhom-bse-detail},
             all building blocks of the nucleon current matrix element in Eq.~\eqref{emcurrent-gauging} are now specified.

             To summarize, the gauging method provides in combination with Eq.~\eqref{G-def} a systematic and intuitive way of computing
             hadronic matrix elements of local quark currents. If a current $\Gamma_0$ probes a hadron, it does so via Eq.~\eqref{emcurrent-gauging},
             i.e., the interaction is resolved in a coupling to the quarks and the kernels (which
             are microscopically again resolved into a coupling to quarks).
             The interaction of the current with the dressed quark is described by Eq.~\eqref{G-def} which amounts to the sum of a pointlike part plus
             all possible reaction mechanisms between quark and antiquark which constitute the T-matrix.
             In practice, the latter part is accessible by resumming the full vertex via solving its inhomogeneous BSE.
             We finally note that Eq.~\eqref{emcurrent-gauging} is valid for arbitrary kernel ans\"atze.
             Thus, if the two- and three-body kernels are known diagrammatically, the method can be applied for implementing
             interactions beyond rainbow-ladder as well.


\section{Axial and pseudoscalar form factors} \label{sec:ffs}

             We continue by discussing the implications of the relations collected in Section~\ref{sec:vertices} for
             the axialvector and pseudoscalar current matrix elements in the nucleon.
             The axialvector current is characterized by two form factors:
             the axial form factor $G_A(Q^2)$ whose value at zero momentum transfer is the nucleon's axial coupling constant $g_A$,
             and the induced pseudoscalar form factor $G_P(Q^2)$. They constitute the nucleon matrix elements of the axialvector current in Eq.~\eqref{currents-general} via
             \begin{equation}\label{ax-current}
                 J^\mu_5(P,Q) = \Lambda_+^f \gamma_5 \left( G_A \gamma^\mu + G_P\,\frac{i  \,Q^\mu}{2M}  \right) \Lambda_+^i\,,
             \end{equation}
             where $P_i$ and $P_f$ are the incoming and outgoing nucleon momenta, $Q = P_f-P_i$ is the momentum transfer,
             $P = (P_i + P_f)/2$ is the average momentum, and $M$ is the nucleon mass.
             A further possible structure proportional to $P^\mu$ is forbidden by charge-conjugation invariance.
             Instead of nucleon spinors we use the positive-energy projectors $\Lambda_+^{f,i} = (\mathds{1}+\hat{\Slash{P}}_{f,i})/2$,   
             so that Eq.~\eqref{ax-current} is a matrix in spinor space.
             Similarly, the pseudoscalar current is parametrized by the pseudoscalar form factor $G_5(Q^2)$:
             \begin{equation}\label{ps-current}
                 J_5(P,Q) = G_5\,\Lambda_+^f  \,i\gamma_5\, \Lambda_+^i\,.
             \end{equation}

             It is instructive to split the axialvector current~\eqref{ax-current}
             into purely transverse and longitudinal contributions with respect to the momentum transfer $Q$.
             Inserting the transverse $\gamma-$matrix from Eq.~\eqref{gammam-transverse},
             together with the relation
             $\Lambda_+^f \gamma_5\,\Slash{Q}\,\Lambda_+^i = -2iM\,\Lambda_+^f \gamma_5\,\Lambda_+^i$, yields:
             \begin{equation}
                 J^\mu_5(P,Q) = \Lambda_+^f  \gamma_5 \left( G_A \gamma_T^\mu  - G_L\, \frac{i \,Q^\mu}{2M \tau} \right) \Lambda_+^i\,,
             \end{equation}
             where $\tau = Q^2/(4M^2)$.
             $G_P$ and $G_L$ are related via
             \begin{equation}\label{GP}
                 G_P = \frac{1}{\tau}\left(G_A-G_L\right),
             \end{equation}
             and analyticity for $G_P$ at vanishing momentum transfer requires $G_L(0)= G_A(0)$.

            The relations of Section~\ref{sec:vertices} directly translate to the nucleon level
            since the form-factor diagrams of Fig.~\ref{fig:current} are identical in
            the axialvector and pseudoscalar cases  except for the type of vertex that is involved.
             Thus, the properties of the $q\bar{q}$ vertices $\Gamma^\mu_{5,T}$, $\Gamma_{5,L}$ and $\Gamma_5$
             are reflected in the form factors $G_A$, $G_L$ and $G_5$, respectively.
             For timelike values $Q^2<0$, the axial form factor $G_A$ must inherit the transverse axial-vector poles
             encompassed in $\Gamma^\mu_{5,T}$, whereas the pseudoscalar pole structure will appear in both $G_L$ and $G_5$.
             Eq.~\eqref{GP} implies that the induced pseudoscalar form factor $G_P$ will inherit both pseudoscalar and
             axial-vector poles.

             On the pion's mass shell, the residue of the pseudoscalar form factor $G_5$ is the pion-nucleon coupling constant
             which can be made explicit by defining
             \begin{equation}\label{GpiNN-def}
                  G_5(Q^2) = \frac{m_\pi^2}{Q^2+m_\pi^2}\,\frac{f_\pi}{m_q}\,G_{\pi NN}(Q^2)\,,
             \end{equation}
             with $G_{\pi NN}(-m_\pi^2) = g_{\pi NN}$. This is the analogue of the vertex relations~\eqref{G-Pole2} and~\eqref{rpi}
             at the form-factor level, and the missing factor $2$ is a consequence of the flavor conventions:
             the $q\bar{q}$ vertices carry flavor matrices $\sigma_k/2$ and the pion's bound-state amplitude a factor $\sigma_k$.

             The AXWTI~\eqref{ax-wti} at the nucleon level becomes
             \begin{equation}\label{axwti-nucleon0}
                 Q^\mu J^\mu_5  + 2 m_q\,J_5= 0
             \end{equation}
             and thereby relates the longitudinal with the pseudoscalar form factor~\cite{Ishii:2000zy,Scherer:2002tk}:
             \begin{equation}\label{axwti-nucleon}
                 G_L = \frac{m_q}{M}\,G_5 = \frac{m_\pi^2}{Q^2+m_\pi^2}\,\frac{f_\pi}{M}\,G_{\pi NN}\,.
             \end{equation}
             Combined with the analyticity requirement for $G_P$ from Eq.~\eqref{GP},
             this yields the Goldberger-Treiman relation:
             \begin{equation}\label{gtr}
                 G_A(0) =  \frac{f_\pi}{M}\,G_{\pi NN}(0)\,,
             \end{equation}
             which is valid for all current-quark masses.
             Thus, the axial and pseudoscalar currents~(\ref{ax-current}--\ref{ps-current})
             are described by two independent form factors $G_A$ and $G_{\pi NN}$ which, at $Q^2=0$, are related via the Goldberger-Treiman relation.

            Implementing the AXWTI and the analyticity constraint explicitly, one can further define
            \begin{equation}\label{R-def}
                G_A(Q^2) = \frac{f_\pi}{M}\,G_{\pi NN}(Q^2) - \frac{Q^2}{4M^2}\, R(Q^2)\,,
            \end{equation}
            so that the quantity $R(Q^2)$ represents the deviation from the Goldberger-Treiman relation at non-zero momentum transfer.
            In that way the two currents are parametrized by two independent form factors $G_A(Q^2)$ and $R(Q^2)$, or $G_{\pi NN}(Q^2)$ and $R(Q^2)$.
            This yields for the induced pseudoscalar form factor via Eqs.~\eqref{GP} and~\eqref{axwti-nucleon}:
            \begin{equation}\label{GP-ppd}
                G_P = \frac{4M^2 G_A - m_\pi^2 R}{Q^2+m_\pi^2} =   \frac{4M f_\pi\,G_{\pi NN} }{Q^2+m_\pi^2}-R\,.
            \end{equation}
            If $G_P$ is expressed through $G_A$, the form factor $R$ enters in combination with $m_\pi^2$;
            if $G_P$ is related to $G_{\pi NN}$, it appears as the non-resonant remainder at the pion pole.
            The pion-pole dominance assumption which corresponds to $R=0$ in Eq.~\eqref{GP-ppd} is therefore justified,
            at least in the low quark-mass region and the neighborhood of the pion pole.
            The contribution from $R$ is however not necessarily small in Eq.~\eqref{R-def} itself
            since that would entail the validity of the Goldberger-Treiman relation for all $Q^2$.

            We finally recall that the 'Goldberger-Treiman discrepancy' in the usual sense is defined as
            \begin{equation}\label{GT-Discrepancy}
                \Delta_\text{GT} = 1 - \frac{G_A(0)}{\frac{f_\pi}{M}\,G_{\pi NN}(-m_\pi^2)}
                                 \stackrel{\eqref{gtr}}{=}  1 - \frac{G_{\pi NN}(0)}{G_{\pi NN}(-m_\pi^2)}
            \end{equation}
            and measures the discrepancy of $G_{\pi NN}$ between $Q^2=0$ and the pion pole at a given current-quark mass,
            and thereby the distance from the chiral limit.

        \begin{figure*}[t]
                    \begin{center}

                    \includegraphics[scale=0.36]{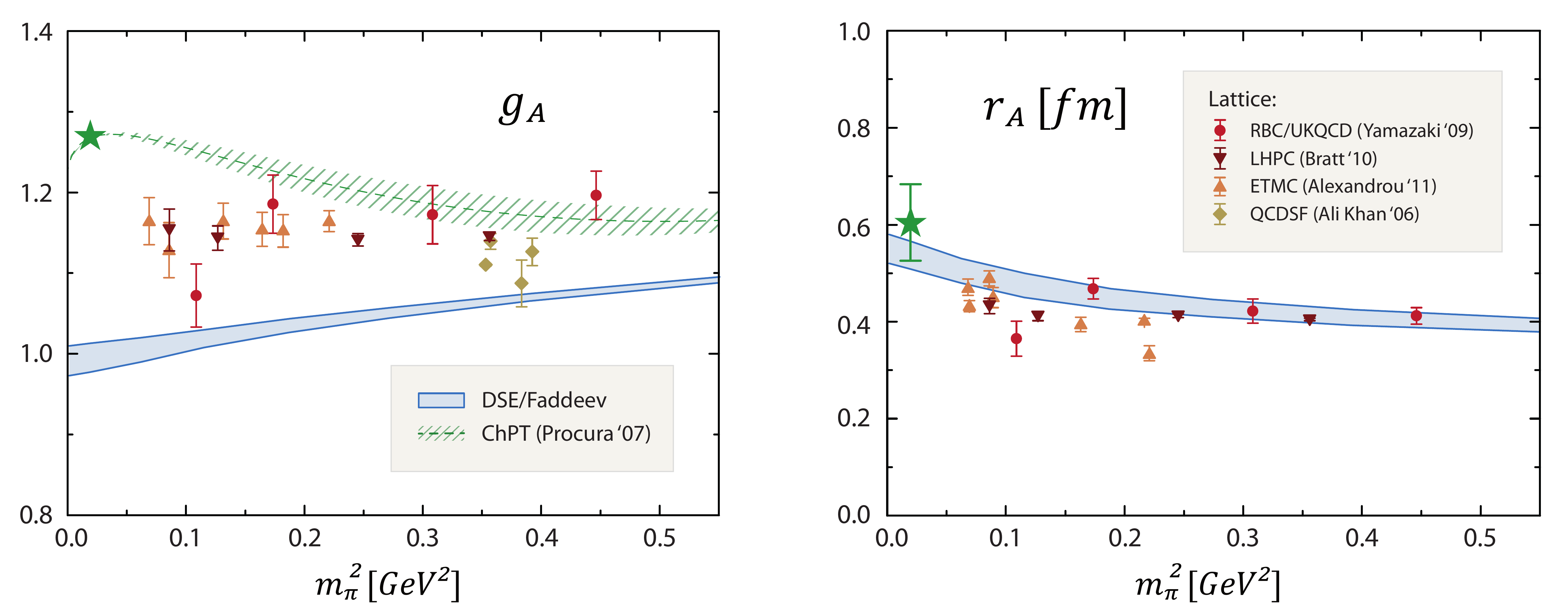}
                    \caption{(Color online) Results for the nucleon's axial charge $g_A$ and axial radius $r_A$ as a function of the pion mass.
                             We compare to lattice results from Refs.~\cite{Khan:2006de,Yamazaki:2009zq,Bratt:2010jn,Alexandrou:2010hf}
                             and the chiral expansion of Ref.~\cite{Procura:2006gq} that includes explicit $\Delta$ degrees of freedom.
                             Stars denote the experimental values.}\label{fig:ga-ra}

                    \end{center}
        \end{figure*}

\section{Results and discussion} \label{sec:results}

          Before discussing the results, let us briefly revisit the successive steps that are performed in the present calculation.
          Once the rainbow-ladder kernel is specified via Eqs.~\eqref{RLkernel} and~\eqref{couplingMT},
          we solve the Dyson-Schwinger equation~\eqref{quarkdse} for the quark propagator, the covariant
          Faddeev equation~\eqref{faddeev:eq} for the nucleon amplitude,
          and the inhomogeneous Bethe-Salpeter equations~(\ref{inhom-bse-detail}--\ref{inhom-bse-detail-2}) for the axialvector and pseudoscalar quark-antiquark vertices.
          These quantities enter the nucleon's current matrix elements~\eqref{emcurrent-gauging} whose decomposition is illustrated in Fig.~\ref{fig:current}.
          The explicit construction of the form-factor integrals is discussed in detail in App.~\ref{sec:emcurrent-worked-out}.
          The axial, longitudinal and pseudoscalar form factors
          $G_A(Q^2)$, $G_L(Q^2)$ and $G_5(Q^2)$ are finally extracted from the Dirac traces in Eq.~\eqref{ff-traces}.

\begin{figure*}[t]
                    \begin{center}

                    \includegraphics[scale=0.36]{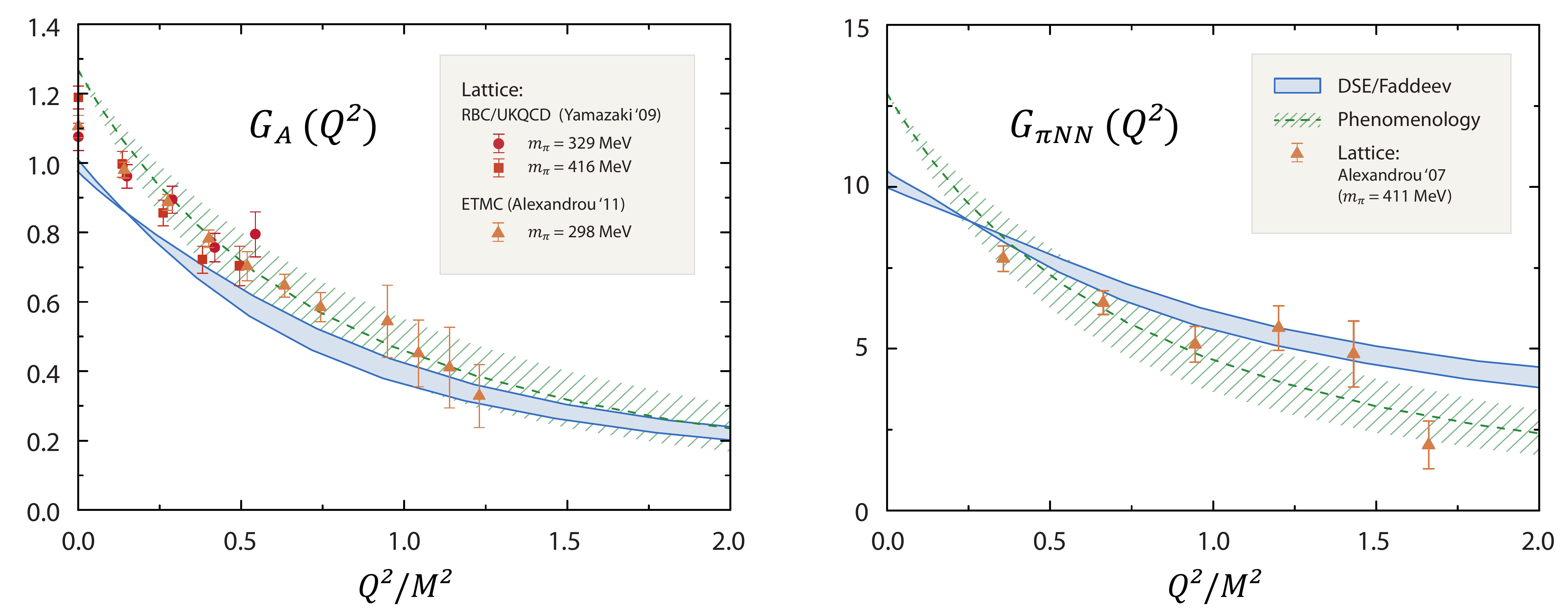}
                    \caption{(Color online) Results for $G_A(Q^2)$ and $G_{\pi NN}(Q^2)$ at the physical $u/d$ mass.
                    We compare to a selection of lattice results at different pion masses~\cite{Bratt:2010jn,Alexandrou:2007xj,Alexandrou:2010hf}.
                    The dashed line with the hatched area in the plot for $G_A$ is the experimental dipole form~\eqref{GA-Dipole} with an axial mass $m_A=1.15 \pm 0.15$ GeV.
                    The phenomenological curve for $G_{\pi NN}(Q^2)$ is obtained from the dipole ansatz for $G_A$
                    by assuming the validity of the Goldberger-Treiman relation for all $Q^2$, i.e., Eq.~\eqref{R-def} with $R=0$. }\label{fig:GA-GPiNN}

                    \end{center}
        \end{figure*}

          Within the rainbow-ladder truncation, the only model input in our calculation is the effective quark-gluon interaction from Eq.~\eqref{couplingMT}.
          It depends on two infrared parameters, the scale $\Lambda$ and the width parameter $\eta$.
          The calculation is carried out with fixed $\Lambda=0.74$~GeV that is chosen to reproduce the experimental pion decay constant at the physical pion mass.
          The width parameter $\eta=1.8 \pm 0.2$ is a measure of the model uncertainty and leads to the colored bands in Figs.~(\ref{fig:ga-ra}--\ref{fig:GP}).
          By modifying the current-quark mass that enters the quark DSE as an input, we can further study the quark-mass dependence of the form factors.
          The quark mass can be mapped onto the pion mass by simultaneously solving the BSE for the pion with the same input.
          The resulting values for the pion decay constant, the nucleon mass, and various other observables related to the
          axial and pseudoscalar currents are tabulated in Table~\ref{tab:results} for three different pion masses.

    \renewcommand{\arraystretch}{1.2}

             \begin{table*}[t]

                \begin{center}
                \begin{tabular}{   c @{\;\;} ||  @{\;\;}c@{\;\;} | @{\;\;}c@{\;\;} | @{\;\;}c@{\;\;} ||
                                                 @{\;\;}c@{\;\;} | @{\;\;}c@{\;\;} | @{\;\;}c@{\;\;} | @{\;\;}c@{\;\;}||
                                                 @{\;\;}c@{\;\;} | @{\;\;}c@{\;\;} | @{\;\;}c@{\;\;}     }

                           &  $m_\pi$  &  $f_\pi$  &  $M_N$      &  $g_A$      &  $m_A$      & $m_A/M_N$     &  $m_{a_1}/M_N$ &  $G_{\pi NN}(0)$ &  $g_{\pi NN}$ & $\Delta_\text{GT}$   \\   \hline

                    Exp.   &  $0.138$  &  $0.092$  &  $0.94$     &  $1.27$     &  $1.0 \dots 1.3$ & $1.1\dots 1.4$ & $1.32(3)$ & $12.9$          &  $13.2$    & $0.02$                     \\  \hline

                           &  $0.138$  &  $0.092$  &  $0.94(1)$  &  $0.99(2)$  &  $1.28(6)$  &  $1.36(7)$    &  $0.95$        &  $10.2(3)$       &  $10.4(3)$    & $0.01$                    \\
                    Calc.  &  $0.522$  &  $0.115$  &  $1.25(2)$  &  $1.05(1)$  &  $1.51(4)$  &  $1.21(5)$    &  $0.89$        &  $11.5(2)$       &  $13.4(5)$    & $0.14(1)$              \\
                           &  $0.761$  &  $0.134$  &  $1.51(3)$  &  $1.10   $  &  $1.69(3)$  &  $1.12(4)$    &  $0.86$        &  $12.6(2)$       &  $16.4(6)$    & $0.23(2)$                \\

                \end{tabular} \caption{Results for various observables, as discussed in the text, at three different pion masses and compared to experiment.
                                       The input scale is chosen such that the pion decay constant $f_\pi$ reproduces the experimental value at $m_\pi=138$~MeV.
                                       The parentheses indicate the dependence on the infrared parameter $\eta$ of Eq.~\eqref{couplingMT}.
                                       $m_\pi$, $f_\pi$, $M_N$ and the axial mass $m_A$ are given in GeV, the remaining quantities are dimensionless.
                                       }\label{tab:results}
                \end{center}

        \end{table*}

          \subsection{Quark-mass dependence of $g_A$}\label{sec:ga}

          Our result for the axial charge $g_A=G_A(0)$ as a function of the squared pion mass is shown in the left panel of Fig.~\ref{fig:ga-ra}.
          Its experimental value $g_A=1.2695(29)$ is precisely known from neutron $\beta$ decay~\cite{Nakamura:2010zzi}.
          Depending on the width parameter $\eta$ in the quark-gluon interaction,
          our calculated result $g_A=0.99(2)$ at the physical $u/d$-quark mass underestimates that value by $20-25\%$.
          It also falls below recent lattice data
          but slowly increases with the pion mass and approaches the chiral expansion from Ref.~\cite{Procura:2006gq} at higher quark masses.

          The determination of $g_A$ in chiral perturbation theory has been notoriously difficult
          due to strong cancelations at leading and next-to-leading order in the chiral expansion~\cite{Bernard:2001rs,Bernard:2006te}.
          An important role in the nucleon's axial structure has been attributed to the $\Delta(1232)$~\cite{Jenkins:1991es,Detmold:2002nf,Hemmert:2003cb}.
          The implementation of pion loops with internal $\Delta$ degrees of freedom in the chiral expansion
          produces indeed a relatively flat pion-mass dependence of $g_A$~\cite{Procura:2006gq} which is plotted in Fig.~\ref{fig:ga-ra} for comparison.
          Lattice results for $g_A$ with pion masses below $300$~MeV are now available~\cite{Yamazaki:2009zq,Bratt:2010jn,Alexandrou:2010hf}
          but have not been able to clarify the issue as they still underestimate the axial charge by $10-15\%$, cf. Fig.~\ref{fig:ga-ra}.
          They are, however, strongly sensitive to finite-volume effects which has been interpreted
          as a signal of missing pion-cloud contributions on the lattice~\cite{Ohta:2011nv}.

          A similar interpretation arises in the Dyson-Schwinger approach.
          From a microscopic point of view, pion exchange corresponds to a gluon resummation.
          While the pion's bound-state pole is recovered in the $q\bar{q}$ scattering matrix in rainbow-ladder truncation,
          which corresponds to a gluon resummation in the $s$-channel,
          pion-cloud contributions are generated by $t-$channel pion exchange~\cite{Fischer:2008wy} whose relevant
          gluon topologies are not captured in rainbow-ladder.
          In fact, missing pion-cloud contributions have turned out to be the main source of discrepancy in the chiral and low-momentum structure
          of the nucleon's electromagnetic form factors computed in the present setup~\cite{Eichmann:2011vu}.
          For example, the nucleon's isoscalar anomalous magnetic moment $\kappa_s$, where leading-order chiral corrections cancel,
          is accurately reproduced by the Faddeev calculation;
          the nucleon charge radii underestimate the experimental values but converge
          with lattice data at larger quark masses; and the low-$Q^2$ behavior of the form factors shows missing structure
          whereas one finds reasonable agreement with experiment at larger $Q^2$.
          These observations suggest to identify the rainbow-ladder truncated nucleon
          with the 'quark core' in chiral effective field theories.
          In the case of axial form factors it would imply
          that the discrepancy between our result for $g_A$ and its experimental value (and its chiral expansion)
          mainly owes to chiral cloud effects as well.

          Irrespective of that, we find that the Goldberger-Treiman relation
          of Eq.~\eqref{gtr} is satisfied quite accurately. The relation
          \begin{equation} \label{gtr-2}
             \frac{f_\pi\, G_{\pi NN}(0)}{M_N\,G_A(0)} =1
          \end{equation}
          is realized at the permille level at the physical $u/d$ mass
          and still reproduced up to $1-2\%$ at the largest quark mass where our numerical treatment becomes increasingly inaccurate.
          We note that a translationally invariant cutoff regularization in the quark DSE and vertex equations~\cite{Holl:2005vu}
          turned out to be crucial in order to ensure its validity;
          a hard numerical integral cutoff leads to violations of several percent in Eqs.~\eqref{analyticity} and \eqref{gtr-2}.
          This procedure also required a slight readjustment of the input scale $\Lambda$ compared to Ref.~\cite{Eichmann:2011vu}
          to reproduce the experimental pion decay constant.

          As we have discussed in Sections~\ref{sec:vertices} and~\ref{sec:ffs}, Eq.~\eqref{gtr-2} is a direct consequence of the underlying dynamics.
          The rainbow-ladder kernel satisfies the AXWTI~\eqref{ax-wti} by construction, and that property translates
          to the quark propagator, the quark-antiquark vertices, the nucleon bound-state amplitude and finally also to the form factor level.
          The Goldberger-Treiman relation entails that the same underestimation that is visible in Fig.~\ref{fig:ga-ra} for the axial charge
           also occurs in $G_{\pi NN}(0)$.
          The experimental value for the pion-nucleon coupling constant is $g_{\pi NN} = 13.2$~\cite{Schroder:2001rc},
          and from the Goldberger-Treiman relation one obtains the value $G_{\pi NN}(0)=12.9$ at vanishing momentum transfer.
          Our result $G_{\pi NN}(0)=10.2(3)$ underestimates that phenomenological value again by $20-25\%$, cf.~Table~\ref{tab:results}.

          \subsection{Momentum dependence of $G_A$}

          While the axial coupling constant is determined from $\beta$ decay,
          experimental data for the momentum dependence of $G_A(Q^2)$ have been obtained in quasielastic neutrino scattering off nucleons or nuclei
          and charged pion electroproduction~\cite{Bernard:2001rs}.
          The existing data can be parametrized by a dipole ansatz,
          \begin{equation}\label{GA-Dipole}
              G_A(Q^2) = \frac{g_A}{\left( 1 + Q^2/m_A^2 \right)^2}\,,
          \end{equation}
          which determines the axial mass $m_A$.
          The axial radius $r_A$ is obtained from the slope of the axial form factor at $Q^2=0$:
          \begin{equation}\label{rA}
              \frac{G_A(Q^2)}{g_A} = 1 - \frac{r_A^2}{6}\,Q^2 + \dots \; \Rightarrow \; r_A^2 = -6\,\frac{G_A'(0)}{g_A}\,.
          \end{equation}
          Inserting the dipole form yields $m_A^2 = 12/r_A^2$, which can also be used to define
          the axial mass independently of the dipole parametrization.

          Due to model-dependent extractions of $G_A$ from the respective cross sections,
          the experimental value of the axial mass is not well constrained.
          Pion electroproduction and older neutrino scattering experiments yield values around $m_A \sim 1$~GeV~\cite{Bernard:2001rs,Lyubushkin:2008pe}
          whereas recent data from MiniBooNE and K2K favor higher central values up to $m_A \sim 1.3$~GeV~\cite{Gran:2006jn,AguilarArevalo:2010zc}.
          The origin of this discrepancy is unclear and could be a consequence of nuclear medium effects~\cite{Nieves:2011yp}
          or a deviation from the dipole form~\cite{Bhattacharya:2011ah}.

          In the right panel of Fig.~\ref{fig:ga-ra} we show our result for the quark-mass dependence of the axial radius $r_A$
          and compare it to lattice data. The corresponding results for the axial mass $m_A=\sqrt{12}/r_A$ are listed in Table~\ref{tab:results}
          for three different pion masses. In contrast to $g_A$, we find reasonable agreement with the lattice data throughout the current-mass range.
          The results are also compatible with experiment at the physical $u/d$ mass as long as a somewhat larger experimental value
          for the axial mass is assumed, i.e., $m_A \gtrsim 1.2$~GeV.

          The $Q^2-$evolution of $G_A$ is depicted in Fig.~\ref{fig:GA-GPiNN} and compared to lattice results and the phenomenological
          dipole parametrization of Eq.~\eqref{GA-Dipole}.
          At low $Q^2$, our result underestimates the lattice data and the dipole curve
          whereas they converge above $Q^2 \sim 1\dots 2$~GeV$^2$.
          This is again a similar behavior as in the case of electromagnetic form factors, and it might indicate
          missing chiral cloud effects in the low-momentum region.
          From that point of view, our calculation favors again a somewhat larger experimental value for the axial dipole mass since for smaller values of $m_A$
          the dipole curve will fall below our result already in the low-$Q^2$ region.

          In general, the low-$Q^2$ behavior of the form factors is dominated by their timelike singularity structure.
          This feature is quite transparent in the Dyson-Schwinger approach where it
          is induced by meson poles in the vertices at timelike $Q^2$ which have their origin in the $q\bar{q}$ scattering matrix, cf.~Eqs.~(\ref{G-pole}--\ref{G-Pole2}).
          While the relevant mass scale in the case of electromagnetic form factors is the $\rho-$meson,
          $G_A$ corresponds to the transverse part of the axialvector $q\bar{q}$ vertex $\Gamma_{5,T}^\mu$
          which has meson poles in the axialvector ($J^{PC}=1^{++}$) isovector channel.
          Thus, the axial mass will be dominated by the properties of the axialvector meson $a_1(1260)$
          and, to some extent, also by its radial excitations.

          While we cannot directly access the form factors for timelike values of $Q^2$ without a complex continuation in the Faddeev amplitude,
          computing the $q\bar{q}$ vertices at timelike momenta is unproblematic.
          From that we can conclude that our result for $G_A$ must exhibit a pole at $Q^2=-m_{a_1}^2$. Thus, if $G_A$ were to follow a perfect dipole form,
          our calculated axial mass would coincide with the result for $m_{a_1}$ obtained in the same setup.
          The mass of the $a_1$ meson is considerably underestimated in rainbow-ladder truncation, with
          $m_{a_1} =0.90(1)$~GeV depending on the parameter $\eta$;
          see Refs.~\cite{Blank:2011qk,Blank:2011ha} for a comprehensive collection of ground-state and excited meson masses in rainbow-ladder.
          This deficiency is well known and corrections beyond rainbow-ladder can be held accountable for it~\cite{Fischer:2009jm,Chang:2011vu}.
          On the other hand, our result for the axial mass extracted from the axial radius~\eqref{rA} is $m_A=1.28(6)$ GeV
          which implies that $G_A$ shows a sizeable deviation from the dipole parametrization.

          It is instructive to revisit the current-mass dependence of $G_A(Q^2)$ in view of these considerations.
          We note that the calculated ratio $m_{a_1}/M_N$ decreases only slowly with the quark mass, cf.~Table~\ref{tab:results}.
          A weak current-mass dependence of dimensionless ratios is realized for various observables in the Dyson-Schwinger approach~\cite{Eichmann:2009zx}
          and can be traced back to the interplay of the two scales, $\Lambda$ and $m_q$, that enter the calculation of hadron properties.
          $\Lambda$ represents the scale of dynamical chiral symmetry breaking and enters the definition of the rainbow-ladder interaction~\eqref{couplingMT}.
          It controls the chiral regime and, with the exception of the pion mass, entails a chiral-limit scaling of hadron masses with $\Lambda$~\cite{Nicmorus:2008vb}.
          When increasing the current-quark mass $m_q$, the impact of dynamical chiral symmetry breaking
          diminishes since it is roughly quark-mass independent, and hadron masses scale with $m_q$.

          A weak dependence of $m_{a_1}/M_N$ on the quark mass implies that the pole position of the $a_1-$meson
          in the variable $Q^2/M^2$, and therefore the overall shape of the form factor $G_A$ as a function of $Q^2/M^2$,
          will be insensitive to a change in the current-quark mass.
          We find indeed that the $Q^2/M^2-$behavior of the form factor $G_A$ in Fig.~\ref{fig:GA-GPiNN}
          does not appreciably change when going to higher quark masses.
          The same can be said about the lattice data of Refs.~\cite{Yamazaki:2009zq,Alexandrou:2010hf}:
          when plotted over $Q^2/M^2$, the spread between the data obtained at various pion masses
          (of which only a subset at the lightest available pion masses are shown in Fig.~\ref{fig:GA-GPiNN})
          is significantly reduced, and they fall at a relatively narrow band
          that agrees well with the dipole parametrization at the physical point.
          Similar properties have been observed for various other form factors, e.g.,
          the electromagnetic form factors of the nucleon~\cite{Eichmann:2011vu} and the $\Delta$ resonance~\cite{Nicmorus:2010sd}.
          Of course, genuine chiral features such as non-analyticities stemming from the pion cloud and the opening of
          decay channels would lead to a deviation from this behavior;
          however, such features are not captured by our present truncation.

        \begin{figure}[t]
                    \begin{center}

                    \includegraphics[scale=0.36]{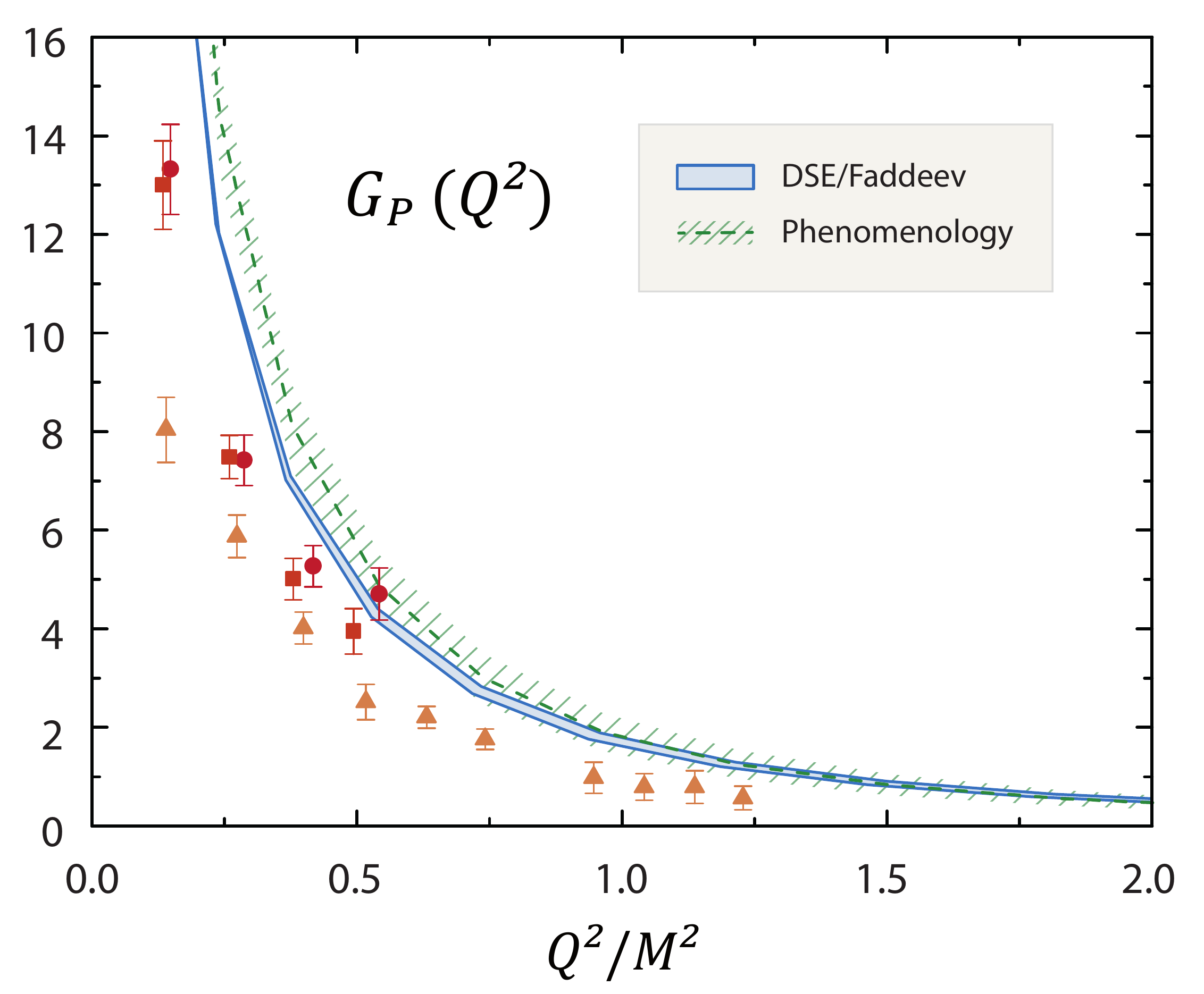}
                    \caption{(Color online) Result for the induced pseudoscalar form factor $G_P(Q^2)$ at the physical $u/d$ mass.
                             The lattice references are the same as in Fig.~\ref{fig:GA-GPiNN} for $G_A(Q^2)$.
                             The phenomenological band corresponds to the pion-pole dominance assumption, i.e., Eq.~\eqref{GP-ppd} with $R=0$,
                             in combination with the dipole parametrization for $G_A$ with an axial mass $m_A=1.15 \pm 0.15$~GeV. }\label{fig:GP}

                    \end{center}
        \end{figure}

          \subsection{Pseudoscalar form factor $G_{\pi NN}$}

          We now turn to the strong pseudoscalar form factor $G_{\pi NN}(Q^2)$
          which is extracted from $G_5(Q^2)$ upon removing the pion pole contribution, cf.~Eq.~\eqref{GpiNN-def}.
          The form factor is not directly observable except at the onshell point $Q^2=-m_\pi^2$,
          where it reproduces the pion-nucleon coupling constant $g_{\pi NN}$.

          The observations from the previous subsection can be carried over here.
          The pole structure in the $0^{-+}$ pseudoscalar vertex that enters the form factor diagrams entails that $G_{\pi NN}$
          is dominated by the properties of the first radial excitation of the pion, i.e., the $\pi(1300)$.
          Its mass in rainbow-ladder truncation is $\sim 1.1$~GeV,
          with a current-mass dependence similar to that of the $a_1$ meson~\cite{Krassnigg:2004if,Krassnigg:2009gd}.
          As a consequence, the form factor $G_{\pi NN}$ as a function of $Q^2/M^2$ shows again a similar behavior throughout the current-mass range.

          In the right panel Fig.~\ref{fig:GA-GPiNN} we show the momentum dependence of $G_{\pi NN}$ at the physical point.
          The phenomenological band denotes the continuation of the Goldberger-Treiman relation~\eqref{gtr} to non-zero $Q^2$ using the dipole parametrization for $G_A$.
          The previous considerations make clear that there is no reason for the Goldberger-Treiman relation, which relates $G_{\pi NN}$ and $G_A$ at $Q^2=0$
          due to analyticity, to hold at non-zero momentum transfer. The properties of both form factors are dominated by different (pseudoscalar and axialvector) meson poles
          with no inherent connection. Nevertheless, one would still expect a qualitatively similar spacelike behavior of the form factors
          since the masses of the $a_1(1260)$ and $\pi(1300)$ are sufficiently close and share a similar quark-mass dependence.

          Fig.~\ref{fig:GA-GPiNN} illustrates that our result for the pseudoscalar form factor deviates indeed from the Goldberger-Treiman relation at $Q^2>0$
          whereas it is compatible with the lattice data of Ref.~\cite{Alexandrou:2007xj}.
          In Table~\ref{tab:results} we also show the result for the pion-nucleon coupling constant $g_{\pi NN}$ at three different quark masses.
          We obtain it by extrapolating $G_{\pi NN}$ from spacelike momenta to $Q^2=-m_\pi^2$.
          Since the pion mass vanishes in the chiral limit and follows the Gell-Mann-Oakes-Renner relation ($m_\pi^2 \sim m_q$) at small quark masses,
          the location of $Q^2=-m_\pi^2$ will shift to larger timelike values upon increasing the quark mass.
          The form factor $G_{\pi NN}(Q^2/M^2)$ depends only weakly on the quark mass, and therefore
          $g_{\pi NN}$ will grow faster with increasing quark mass than the static value $G_{\pi NN}(0)$;
          see Ref.~\cite{Mader:2011zf} for an analogous discussion of the $\Delta N\pi$ coupling constant.
          As a consequence, the Goldberger-Treiman discrepancy from Eq.~\eqref{GT-Discrepancy},
          which measures the relative difference in $G_{\pi NN}$ between $Q^2=0$ and $Q^2=-m_\pi^2$,
          rises with the squared pion mass, see Table~\ref{tab:results}.

          \subsection{Induced pseudoscalar form factor $G_P$}

          We conclude our analysis with the induced pseudoscalar form factor $G_P(Q^2)$ that enters the axial current of Eq.~\eqref{ax-current}.
          Experimental data for $G_P$ are sparse: its $Q^2-$dependence can be extracted from pion electroproduction, whereas
          ordinary muon capture on the proton ($\mu^-+ p \rightarrow \nu_\mu + n$) determines
          the induced pseudoscalar coupling constant $g_p$:
          \begin{equation}
              g_p = \frac{M_\mu}{2M_N}\,G_P(Q^2=0.88 M_\mu^2)\,,
          \end{equation}
          where $M_\mu$ is the muon mass.
          The recent value reported by the MuCap Collaboration, $g_p = 7.3 \pm 1.1$~\cite{Andreev:2007wg}, is consistent with
          chiral perturbation theory but smaller than the previous world average~\cite{Bernard:2001rs,Gorringe:2002xx}.

          From Eqs.~\eqref{GP} and~\eqref{GP-ppd} it is clear that $G_P(Q^2)$ is dominated by the pion pole at $Q^2=-m_\pi^2$.
          This implies that the pion-pole dominance assumption
          \begin{equation}
              G_P \approx \frac{4M_N^2 \,G_A}{Q^2+m_\pi^2}
          \end{equation}
          describes the spacelike properties of the induced pseudoscalar form factor reasonably well.
          Such a behavior is clearly visible in our result shown in Fig.~\ref{fig:GP}.
          In particular, our underestimation of the axial charge $g_A$ translates to the induced pseudoscalar
          coupling constant and yields $g_P = 6.7(1)$.
          Since $G_P$ is the linear combination of the axial and longitudinal form factors $G_A$ and $G_L$ via Eq.~\eqref{GP},
          we can also infer information about its structure beyond the pion pole: namely, it
          must include all further pseudoscalar ($0^{-+}$) and axialvector ($1^{++}$) pole structures as well.

          In principle, the longitudinal form factor $G_L$ allows to check the validity of the AXWTI~\eqref{axwti-nucleon}
          that relates $G_L$ with the pion-nucleon form factor $G_{\pi N N}$.
          Unfortunately we find that $G_L$ is rather sensitive to the angular dependencies in the nucleon's Faddeev amplitude.
          In the present numerical treatment, the nucleon amplitude is obtained from solving its Faddeev equation in the rest frame.
          In the form factor calculation, a 'Lorentz boost' to a moving frame
          amounts to a complex continuation of the amplitude's dressing functions in its angular variables
          which is usually implemented via Chebyshev expansions.
          Potential convergence problems with growing $Q^2$ can be alleviated by imposing a $Q^2$-dependent cutoff in the Chebyshev expansion~\cite{Eichmann:2011vu}.
           Even then, however, we find that the axial and pseudoscalar form factors at larger $Q^2$ are more sensitive to the numerics than their electromagnetic counterparts.
           This is especially true, even at low momentum transfer, for the longitudinal form factor $G_L(Q^2)$ which is numerically not well under control.
           While that numerical instability does not affect the behavior of the induced pseudoscalar form factor which is dominated by $G_A$ and the pion pole,
           it impedes a test of the AXWTI.

           We want to emphasize that the problem could be entirely eliminated by solving the Faddeev equation in each boosted frame (i.e., at each $Q^2$) anew.
           That would allow for a clean determination of all three form factors $G_A$, $G_P$ and $G_{\pi NN}$ up to $Q^2 \sim 7 \dots 9$~GeV$^2$
           without the need for analytic continuation in the angular variables. The upper $Q^2$ limit is due to the singularities in the quark propagators
           that appear in the integrands; it could be circumvented as well via residue calculus.
           However, these strategies are both numerically and conceptually challenging and beyond the scope of the present work.


\section{Conclusions}\label{sec:conclusions}

      We presented a calculation of the nucleon's axial and pseudoscalar form factors in the framework of Dyson-Schwinger
      and covariant Faddeev equations. We detailed the construction of the axial and pseudoscalar
      currents as well as their microscopic building blocks. 
      All ingredients were computed selfconsistently within a rainbow-ladder truncation which describes the binding of the nucleon
      through iterated dressed gluon exchange between the quarks.
      Since pion-cloud effects are not implemented, our results represent the nucleon's quark core.

      We analyzed the properties of the axial ($G_A$), induced pseudoscalar ($G_P$) and pion-nucleon form factor ($G_{\pi NN}$) from the underlying quark-antiquark vertices
      on which they depend. The spacelike structure of the form factors is connected to the properties of axialvector and pseudoscalar-meson poles in the timelike region.
      $G_A$ is dominated by the axialvector meson $a_1(1260)$ and its radial excitations;
      $G_P$ is governed by the pion pole, and $G_{\pi NN}$ is related to the $\pi(1300)$.
      The Goldberger-Treiman relation follows as a consequence of the axialvector Ward-Takahashi identity and analyticity
      which are both satisfied at the quark-gluon level. There is however no obvious inherent connection between
      the form factors $G_A$ and $G_{\pi NN}$ at non-zero momentum transfer.

      The form factor results at $Q^2=0$ underestimate the experimental values by $20-25\%$ whereas
      they are consistent with the phenomenological dipole form for $G_A$ and the pion-pole dominance ansatz for $G_P$ at larger $Q^2$.
      The nucleon's axial charge $g_A$ falls below the lattice data in
      the low quark-mass region and approaches the result from chiral effective field theory at larger pion masses.
      These features might be signals of missing pion-cloud effects in the chiral and low-momentum regions.

      We encountered difficulties in reproducing the axialvector Ward-Takahashi identity beyond small $Q^2$
      as it turned out to be quite sensitive to the numerical details of the Faddeev amplitude.
      A clean determination of the form factors at larger $Q^2$ would require a moving-frame solution of the Faddeev equation.
      This is numerically challenging and remains a task for future investigations.
      Moreover, in order to decide whether the missing structure in the form factors is indeed attributable to pion-cloud effects,
      a consistent description of baryons beyond the rainbow-ladder truncation is necessary.
      Steps in that direction are planned.


     \section{Acknowledgements}

     We are grateful to R. Alkofer, M. Blank, and A. Krassnigg for valuable discussions.
     This work was supported by the Austrian Science Fund FWF under
     Erwin-Schr\"odinger-Stipendium No.~J3039,
            the Helmholtz International Center for FAIR
            within the LOEWE program of the State of Hesse,
            and the Helmholtz Young Investigator Group No.~VH-NG-332.


\begin{appendix}

  \section{Euclidean conventions} \label{app:conventions}

            We work in Euclidean momentum space with the following conventions:
            \begin{equation}
                p\cdot q = \sum_{k=1}^4 p_k \, q_k,\quad
                p^2 = p\cdot p,\quad
                \Slash{p} = p\cdot\gamma\,.
            \end{equation}
            A vector $p$ is spacelike if $p^2 > 0$ and timelike if $p^2<0$.
            The hermitian $\gamma-$matrices $\gamma^\mu = (\gamma^\mu)^\dag$ satisfy the anticommutation relations
            $\left\{ \gamma^\mu, \gamma^\nu \right\} = 2\,\delta^{\,\mu\nu}$, and we define
            \begin{equation}
                \sigma^{\mu\nu} = -\frac{i}{2} \left[ \gamma^\mu, \gamma^\nu \right]\,, \quad
                \gamma^5 = -\gamma^1 \gamma^2 \gamma^3 \gamma^4\,.
            \end{equation}
            In the standard representation one has:
            \begin{equation*}
                \gamma^k  =  \left( \begin{array}{cc} 0 & -i \sigma_k \\ i \sigma_k & 0 \end{array} \right), \;
                \gamma^4  =  \left( \begin{array}{c@{\quad}c} \mathds{1} & 0 \\ 0 & \!\!-\mathds{1} \end{array} \right), \;
                \gamma^5  =  \left( \begin{array}{c@{\quad}c} 0 & \mathds{1} \\ \mathds{1} & 0 \end{array} \right),
            \end{equation*}
            where $\sigma_k$ are the three Pauli matrices.
            The charge conjugation matrix is given by
            \begin{equation}
                C = \gamma^4 \gamma^2, \quad C^T = C^\dag = C^{-1} = -C\,,
            \end{equation}
            and the charge conjugates for (pseudo-)\,scalar, \mbox{(axial-)} vector and tensor amplitudes are defined as
            \begin{equation}\label{chargeconjugation}
            \begin{split}
                \conjg{\Gamma}(p,P) &:= C\,\Gamma(-p,-P)^T\,C^T \,,   \\
                \conjg{\Gamma}^\alpha(p,P) &:= -C\,{\Gamma^\alpha}(-p,-P)^T\,C^T \,,   \\
                \conjg{\Gamma}^{\beta\alpha}(p,P) &:= C\,{\Gamma^{\alpha\beta}}(-p,-P)^T\,C^T\,,
            \end{split}
            \end{equation}
            where $T$ denotes a Dirac transpose.
            Four-momenta are conveniently expressed through hyperspherical coordinates:
            \begin{equation}\label{APP:momentum-coordinates}
                p^\mu = \sqrt{p^2} \left( \begin{array}{l} \sqrt{1-z^2}\,\sqrt{1-y^2}\,\sin{\phi} \\
                                                           \sqrt{1-z^2}\,\sqrt{1-y^2}\,\cos{\phi} \\
                                                           \sqrt{1-z^2}\;\;y \\
                                                           \;\; z
                                         \end{array}\right),
            \end{equation}
            and a four-momentum integration reads:
            \begin{equation*} \label{hypersphericalintegral}
                 \int\limits_p := \frac{1}{(2\pi)^4}\,\frac{1}{2}\int\limits_0^{\infty} dp^2 \,p^2 \int\limits_{-1}^1 dz\,\sqrt{1-z^2}  \int\limits_{-1}^1 dy \int\limits_0^{2\pi} d\phi \,.
            \end{equation*}
            We frequently refer to quantities that are 'transverse' with respect to a four-momentum $Q$.
            The precise meaning of that is expressed by the transverse projector
            \begin{equation}
               T_Q^{\mu\nu} = \delta^{\mu\nu} - \hat{Q}^\mu \hat{Q}^\nu \,,
            \end{equation}
            where the normalized momentum is $\hat{Q}^\mu = Q^\mu/\sqrt{Q^2}$, so that $\gamma-$matrices that are transverse to $Q$ read:
            \begin{equation}
                \gamma_T^\mu = T_Q^{\mu\nu} \gamma^\nu = \gamma^\mu - \hat{Q}^\mu \hat{\Slash{Q}} \,.
            \end{equation}
            They satisfy $\left\{ \gamma_T^\mu, \gamma_T^\nu \right\} = 2\,T_Q^{\mu\nu}$ and $\gamma_T^\mu\, \gamma_T^\mu =3$.
            Similarly, momenta transverse to $Q$ are defined by
            \begin{equation}
                p_T^\mu = T_Q^{\mu\nu} p^\nu = p^\mu - (p\cdot\hat{Q})\,\hat{Q}^\mu\,.
            \end{equation}

            \begin{figure*}[t]
            \begin{center}
            \includegraphics[scale=0.178]{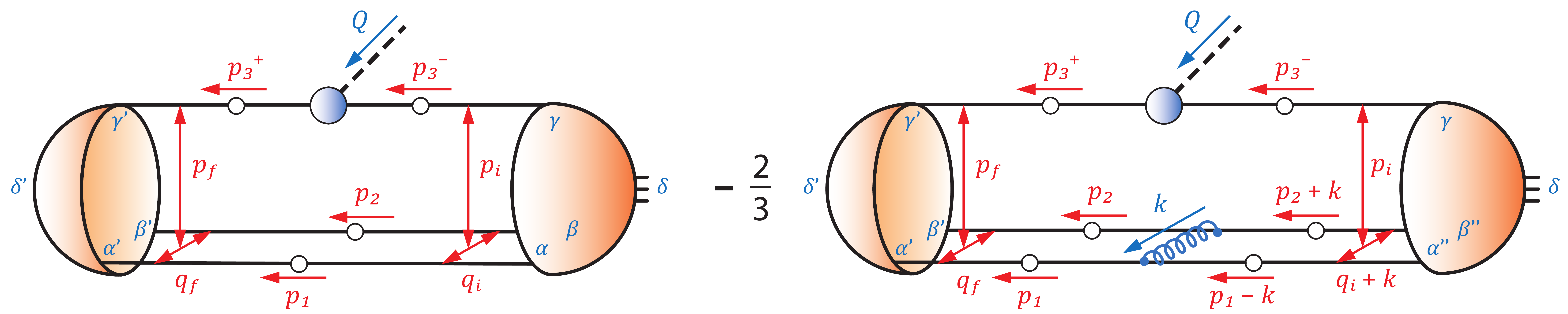}
            \caption{(Color online)
                     Notation and kinematics in the nucleon's three-body current in rainbow-ladder truncation.} \label{fig:current-kinematics}
            \end{center}
            \end{figure*}

    \section{Current diagrams in the three-quark framework} \label{sec:emcurrent-worked-out}

 \renewcommand{\arraystretch}{1.6}

            In this appendix we detail the construction of the nucleon's axialvector and pseudoscalar current
            in rainbow-ladder truncation which is illustrated in Figs.~\ref{fig:current} and~\ref{fig:current-kinematics}.
            We first collect the properties of the nucleon's covariant bound-state amplitude that enters the current diagrams.
            The nucleon amplitude including its full Dirac, flavor and color dependence reads
              \begin{equation}\label{FE:nucleon_amplitude_full}
                  \mathbf{\Psi}(p,q,P) = \left( \sum_{n=1}^2  \Psi_n \, \mathsf{F}_n\right) \frac{\varepsilon_{ABC}}{\sqrt{6}}\,,
              \end{equation}
            where $[\Psi_n]_{\alpha\beta\gamma\delta}(p,q,P)$ is the spin-momentum amplitude and $[\mathsf{F}_n]_{abcd}$
            are the flavor tensors. Both of them transform as doublets under
            the permutation group $\mathds{S}^3$ (hence the indices $n=1,2$), such that the bracket in Eq.~\eqref{FE:nucleon_amplitude_full} is a permutation-group singlet.
            The Dirac amplitudes $\Psi_n$ carry three spinor indices $\alpha,\beta,\gamma$ for the quark legs and one
            spinor index $\delta$ for the nucleon, and they are mixed-antisymmetric $(\Psi_1)$ or mixed-symmetric $(\Psi_2)$
            under exchange of the indices $\alpha$, $\beta$ and related quark momenta.
            They can be decomposed in a basis of 64 orthonormal Dirac structures $\mathsf{X}^i_{\alpha\beta\gamma\delta}$
            which correspond to $s-$, $p-$ and $d-$waves in the nucleon's rest frame:
            \begin{equation}\label{amplitude:reconstruction1}
                [\Psi_n]_{\alpha\beta\gamma\delta}(p,q,P) = \sum_{i=1}^{64} f_n^i(t)\,\mathsf{X}^i_{\alpha\beta\gamma\delta}(p,q,P)\,,
            \end{equation}
            see Ref.~\cite{Eichmann:2011vu} for details.
            The Lorentz-invariant dressing functions $f_n^i(t)$ are the solutions of the covariant Faddeev equation,
            and $t$ abbreviates the five Lorentz-invariant momentum
            variables that can be constructed from the momenta $p$, $q$ and $P$ ($P^2=-M^2$ is fixed).
            The dressing functions for $n=1,2$ are not independent but related via permutation-group symmetry.

            Similarly, the two isospin-$\nicefrac{1}{2}$ flavor tensors $\mathsf{F}_n$ carry three isospin indices
            $\mathsf{a,b,c}$ for the quarks and one for the nucleon. They are mixed-antisymmetric ($\mathsf{F}_1$) or mixed-symmetric ($\mathsf{F}_2$)
            with respect to $\mathsf{a}$, $\mathsf{b}$:
              \begin{equation} \label{FAD:flavor}
              \begin{split}
                  [\mathsf{F}_1]_\mathsf{abcd} &= \textstyle\frac{1}{\sqrt{2}}\,[i \sigma_2]_\mathsf{ab} \,\delta_\mathsf{cd} \,,\\
                  [\mathsf{F}_2]_\mathsf{abcd} &= -\textstyle\frac{1}{\sqrt{6}}\, [\vect{\sigma}\,i\sigma_2]_\mathsf{ab} \,\vect{\sigma}_\mathsf{cd} \,,
              \end{split}
              \end{equation}
              where the $\sigma_k$ are the Pauli matrices,
             and they are normalized to unity: $[\mathsf{F}^\dag_{n'}]_{\mathsf{bad'c}}[\mathsf{F}_n]_{\mathsf{abcd}} = \delta_{n'n} \,\delta_{\mathsf{d'd}}$.
             To project onto proton or neutron flavor states, the nucleon index $\mathsf{d}$ must be contracted with either of the two isospin vectors $(1,0)$ or $(0,1)$, respectively.
             Finally, the antisymmetric tensor $\varepsilon_{ABC}$ in Eq.~\eqref{FE:nucleon_amplitude_full} denotes the color part which we also normalize to $1$.

            The rainbow-ladder truncated current is the sum of the impulse-approximation and kernel diagrams of Fig.~\ref{fig:current-kinematics}.
            In principle one must also add up all three permutations $a=1,2,3$, where $a$ is the label of the quark that couples to the current.
            However, the symmetry properties of the nucleon amplitude relate the permuted form-factor diagrams among each other via Eqs.~\eqref{current:permutations} below, such that
            it is sufficient to consider only one of the three permutations explicitly. In the following we choose $a=3$,
            which corresponds to a coupling to the upper quark line as shown in Fig.~\ref{fig:current-kinematics}.
            We discuss the construction of the axialvector current $J^\mu_5$; the equations
            for the pseudoscalar current $J_5$ are completely analogous.

            Starting with the flavor traces, the axialvector and pseudoscalar $q\bar{q}$ vertices
             come with flavor factors $\sigma_k/2$ whereas the rainbow-ladder kernel is flavor-independent. Hence,
             the flavor trace in both $a=3$ diagrams is given by
             \begin{equation} \label{flavor-current}
                 \left[\mathsf{F}^{(3)}_{n'n}\right]^k_\mathsf{d'd} := [\mathsf{F}^\dag_{n'}]_{\mathsf{bad'c'}} \left[\frac{\sigma_k}{2}\right]_{\mathsf{c'c}} \,[\mathsf{F}_n]_{\mathsf{abcd}}\,.
             \end{equation}
             We keep the doublet indices $n'$ and $n$ for the outgoing and incoming nucleon amplitude general
             so we can treat the Dirac and flavor parts in the current separately.
             The above expression vanishes if $n\neq n'$, and for $n=n'$ it yields:
             \begin{equation}
                 \left[\mathsf{F}^{(3)}_{11}\right]^k = -3\left[\mathsf{F}^{(3)}_{22}\right]^k  = \frac{\sigma_k}{2}\,.
             \end{equation}
            Next, the color trace in the impulse-approximation diagram equals $1$ while the kernel diagram picks up
            a color factor $-\nicefrac{2}{3}$.
            The current for the coupling to quark $a=3$, which is a matrix in Dirac and flavor space, can then be written as
            \begin{equation}\label{current-3}
                \left[J_{n'n}^{(3)}\right]^{\mu} =   \left[J^{(3),\text{IMP}}_{n'n} - \textstyle\frac{2}{3} \,J^{(3),\text{K}}_{n'n}\right]^\mu_{\delta'\delta}
                \left[\mathsf{F}^{(3)}_{n'n}\right]^k_\mathsf{d'd}\,,
            \end{equation}
            where the first bracket includes the Dirac parts. We suppressed the Dirac and isospin indices
            on the left-hand side and the index $\mu$ only appears in the axialvector case. The total current of Eq.~\eqref{ax-current}
            is then the sum over the three permutations and the final and initial doublet configurations:
             \begin{equation}\label{current-total}
                 J_5^{\mu}(P,Q) = \sum_{a=1}^3 \sum_{n'n} \left[J_{n'n}^{(a)}\right]^{\mu}\,.
             \end{equation}

            Let us now examine the Dirac contributions in Eq.~\eqref{current-3} in more detail.
            The impulse-approximation diagram reads explicitly:
             \begin{equation}\label{current:diagram-3}
             \begin{split}
                 &\left[ J^{(3),\text{IMP}}_{n'n} \right]^\mu_{\delta'\delta} =  \int\limits_p \!\!\!\int\limits_q [\conjg{\Psi}_{n'}]_{\beta'\alpha'\delta'\gamma'}(p_f,q_f,P_f) \,\times \\
                 & \times S_{\alpha'\alpha}(p_1)\,S_{\beta'\beta}(p_2)\,\left[ S(p_3^+)\,\Gamma^\mu_5(p_3,Q)\,S(p_3^-)\right]_{\gamma'\gamma} \times \\
                 & \times [\Psi_n]_{\alpha\beta\gamma\delta}(p_i,q_i,P_i)  \,,
             \end{split}
             \end{equation}
             where $p_i$, $q_i$ and $p_f$, $q_f$ are the incoming and outgoing relative momenta;
             $P_i$ and $P_f$ are the incoming and outgoing nucleon momenta;
             $p_1$, $p_2$, $p_3$ and $p_3^\pm=p_3 \pm Q/2$ are the quark momenta;
             $S(p_k)$ are the dressed-quark propagators;
             $\Gamma^\mu_5$ is the dressed axialvector (or, by analogy, pseudoscalar) vertex; and $p$ and $q$ are the two loop momenta.
             For a symmetric momentum-partitioning parameter $\nicefrac{1}{3}$ the relative momenta are explicitly given by
             \begin{equation}\label{ff-momenta-1}
                 p_f = p + \frac{Q}{3}\,, \quad
                 p_i = p - \frac{Q}{3}\,, \quad
                 q_f = q_i = q
             \end{equation}
             and the quark momenta by
             \begin{equation}\label{ff-momenta-2}
                 p_1 = -q - \frac{p}{2} + \frac{P}{3}\,, \quad
                 p_2 =  q - \frac{p}{2} + \frac{P}{3}\,,
             \end{equation}
             and $p_3 = p+P/3$.

             The second diagram involving the rainbow-ladder kernel is identical to Eq.~\eqref{current:diagram-3}
             except that the incoming amplitude $\Psi_n$ in the third row is replaced by $\Psi_n^{(3)}$:
             \begin{equation}\label{current:diagram-kernel-3}
                 \left[ J^{(3),\text{K}}_{n'n} \right]^\mu_{\delta'\delta} = \big[\text{Eq.\,\eqref{current:diagram-3}}\big]_{\Psi_n \rightarrow \Psi_n^{(3)}}\,,
             \end{equation}
             and the latter is defined by
             \begin{equation}\label{current-psi-3}
             \begin{split}
                 &[\Psi_n^{(3)}]_{\alpha\beta\gamma\delta}(p_i,q_i,P_i) =  \int\limits_k K_{\alpha\alpha'\beta\beta'}(k)\,\times \\
                 & \quad\qquad \times S_{\alpha'\alpha''}(p_1-k)\, S_{\beta'\beta''}(p_2+k) \, \times \\
                 & \quad\qquad \times [\Psi_n]_{\alpha''\beta''\gamma\delta}(p_i,q_i+k,P_i) \,.
             \end{split}
             \end{equation}
             Since Eq.~\eqref{current-psi-3} is one of the three diagrams in the Faddeev equation and
             has the same structure as Eq.~\eqref{FE:nucleon_amplitude_full}, it is actually not necessary to compute
             it anew in the form-factor calculation.
             Using the labels $f$ and $i$ to abbreviate the final and initial amplitudes,
             the Dirac bracket in Eq.~\eqref{current-3} can be schematically written as
             \begin{equation*}
             \begin{split}
                 & \int\!\!\!\int \conjg{\Psi}_f\,S_1 \,S_2  \left[S_3 \,\Gamma^\mu_5 \,S_3\right]\left[\Psi_i -\textstyle\frac{2}{3}\,\Psi_i^{(3)}\right]  \\
                         &= \int\!\!\!\int \conjg{\Psi}_f \left[S_3 \,\Gamma^\mu_5\right]\left[S_1 \,S_2 \,S_3\right]\left[ \Psi_i -\textstyle\frac{2}{3}\,\Psi_i^{(3)}\right]   \\
                         &=\int\limits_p \left[S_3\,\Gamma^\mu_5\right] \int\limits_q \conjg{\Psi}_f \left[\Phi_i-\textstyle\frac{2}{3}\,\Phi_i^{(3)}\right],
             \end{split}
             \end{equation*}
             where $\Phi_i$ is the wave function $S_1 S_2 S_3\,\Psi_i$, and $\Phi_i^{(3)}$ is the wave function obtained
             from $\Psi_i^{(3)}$. Both of them can be collected beforehand when solving the Faddeev equation
             and implemented in the form factor diagram simply through evaluation at the proper incoming momenta. Then,
             only the product $S_3 \,\Gamma^\mu_5$, which is independent of the loop momentum~$q$, needs to be computed explicitly.

 \renewcommand{\arraystretch}{1.1}

             Finally, in order to obtain the full current in Eq.~\eqref{current-total}, we need to include the permuted diagrams for $a=1,2$.
             They can be inferred from the $a=3$ diagram by applying the two-dimensional matrix representations $\mathcal{M}'$, $\mathcal{M}''$
             of the permutation group $\mathds{S}^3$,
             \begin{equation*}
                   \mathcal{M}' = \frac{1}{2}\left( \begin{array}{cc} -1 & -\sqrt{3} \\ \sqrt{3} & -1 \end{array}\right), \quad
                   \mathcal{M}'' = \frac{1}{2}\left( \begin{array}{cc} -1 & \sqrt{3} \\ -\sqrt{3} & -1 \end{array}\right)
             \end{equation*}
             which act upon the doublet indices $n',n$ via~\cite{Eichmann:2011vu}
             \begin{equation}\label{current:permutations}
             \begin{split}
                 J^{(1)}_{n'n} &= \left[ \mathcal{M}' J^{(3)} {\mathcal{M}'}^T \right]_{n'n} \left[ \mathcal{M}' \mathsf{F}^{(3)} {\mathcal{M}'}^T \right]_{n'n} \,, \\
                 J^{(2)}_{n'n} &= \left[ \mathcal{M}'' J^{(3)} {\mathcal{M}''}^T \right]_{n'n} \left[ \mathcal{M}'' \mathsf{F}^{(3)} {\mathcal{M}''}^T \right]_{n'n} \,.
             \end{split}
             \end{equation}
             Putting the pieces together and adding up Eq.~\eqref{current:permutations} to obtain~\eqref{current-total} yields the following result for the total isovector current:
             \begin{equation}\label{current-finalfinal}
                 J_5^{\mu}(P,Q) = \left[ 3 J_{11}^{(3)} - J_{22}^{(3)} \right]^\mu_{\delta'\delta} \left[\frac{\sigma_k}{2}\right]_{\mathsf{d'd}}\,,
             \end{equation}
             where $J_{nn}^{(3)}$ are the Dirac contributions to the $a=3$ diagrams, i.e.:
             \begin{equation}
                 J_{nn}^{(3)} = J^{(3),\text{IMP}}_{nn} - \textstyle\frac{2}{3}\,J^{(3),\text{K}}_{nn}\,.
             \end{equation}

             In the final step we want to extract the form factors $G_A$, $G_L$ and $G_5$ that appear in Eqs.~(\ref{ax-current}--\ref{ps-current})
             from the axialvector current $J_5^\mu$ in Eq.~\eqref{current-finalfinal} and its pseudoscalar analogue $J_5$.
             The flavor factors $\sigma_k/2$ on both sides of the equations are the same and can be factored out.
             The form factors are then obtained from the Dirac traces
             \begin{equation}\label{ff-traces}
             \begin{split}
                 G_A &= -\frac{1}{4\,(1+\tau)}\,\text{Tr}\left\{ J_5^\mu \,\gamma_5 \gamma_T^\mu\right\},\\
                 G_L &= \frac{1}{2i\sqrt{\tau}}\,\text{Tr}\left\{ \hat{Q}^\mu J_5^\mu \,\gamma_5\right\},\\
                 G_5  &= \frac{i}{2\tau}\,\text{Tr}\left\{ J_5 \,\gamma_5\right\},
             \end{split}
             \end{equation}
             where $\hat{Q}$ is the normalized momentum transfer, $\gamma_T^\mu$ is transverse to $Q$, and $\tau=Q^2/(4M^2)$.

\end{appendix}


\bigskip

\bibliographystyle{apsrev4-1-mod}

\bibliography{lit-axial}

\begin{thebibliography}{69}%
\makeatletter
\providecommand \@ifxundefined [1]{%
 \@ifx{#1\undefined}
}%
\providecommand \@ifnum [1]{%
 \ifnum #1\expandafter \@firstoftwo
 \else \expandafter \@secondoftwo
 \fi
}%
\providecommand \@ifx [1]{%
 \ifx #1\expandafter \@firstoftwo
 \else \expandafter \@secondoftwo
 \fi
}%
\providecommand \natexlab [1]{#1}%
\providecommand \enquote  [1]{``#1''}%
\providecommand \bibnamefont  [1]{#1}%
\providecommand \bibfnamefont [1]{#1}%
\providecommand \citenamefont [1]{#1}%
\providecommand \href@noop [0]{\@secondoftwo}%
\providecommand \href [0]{\begingroup \@sanitize@url \@href}%
\providecommand \@href[1]{\@@startlink{#1}\@@href}%
\providecommand \@@href[1]{\endgroup#1\@@endlink}%
\providecommand \@sanitize@url [0]{\catcode `\\12\catcode `\$12\catcode
  `\&12\catcode `\#12\catcode `\^12\catcode `\_12\catcode `\%12\relax}%
\providecommand \@@startlink[1]{}%
\providecommand \@@endlink[0]{}%
\providecommand \url  [0]{\begingroup\@sanitize@url \@url }%
\providecommand \@url [1]{\endgroup\@href {#1}{\urlprefix }}%
\providecommand \urlprefix  [0]{URL }%
\providecommand \Eprint [0]{\href }%
\providecommand \doibase [0]{http://dx.doi.org/}%
\providecommand \selectlanguage [0]{\@gobble}%
\providecommand \bibinfo  [0]{\@secondoftwo}%
\providecommand \bibfield  [0]{\@secondoftwo}%
\providecommand \translation [1]{[#1]}%
\providecommand \BibitemOpen [0]{}%
\providecommand \bibitemStop [0]{}%
\providecommand \bibitemNoStop [0]{.\EOS\space}%
\providecommand \EOS [0]{\spacefactor3000\relax}%
\providecommand \BibitemShut  [1]{\csname bibitem#1\endcsname}%
\let\auto@bib@innerbib\@empty
\bibitem [{\citenamefont {Bernard}\ \emph {et~al.}(2002)\citenamefont
  {Bernard}, \citenamefont {Elouadrhiri},\ and\ \citenamefont
  {Meissner}}]{Bernard:2001rs}%
  \BibitemOpen
  \bibfield  {author} {\bibinfo {author} {\bibfnamefont {V.}~\bibnamefont
  {Bernard}}, \bibinfo {author} {\bibfnamefont {L.}~\bibnamefont
  {Elouadrhiri}}, \ and\ \bibinfo {author} {\bibfnamefont {U.~G.}\ \bibnamefont
  {Meissner}},\ }\href {\doibase 10.1088/0954-3899/28/1/201} {\bibfield
  {journal} {\bibinfo  {journal} {J. Phys.}\ }\textbf {\bibinfo {volume}
  {G28}},\ \bibinfo {pages} {R1} (\bibinfo {year} {2002})}\BibitemShut
  {NoStop}%
\bibitem [{\citenamefont {Gorringe}\ and\ \citenamefont
  {Fearing}(2004)}]{Gorringe:2002xx}%
  \BibitemOpen
  \bibfield  {author} {\bibinfo {author} {\bibfnamefont {T.}~\bibnamefont
  {Gorringe}}\ and\ \bibinfo {author} {\bibfnamefont {H.~W.}\ \bibnamefont
  {Fearing}},\ }\href {\doibase 10.1103/RevModPhys.76.31} {\bibfield  {journal}
  {\bibinfo  {journal} {Rev. Mod. Phys.}\ }\textbf {\bibinfo {volume} {76}},\
  \bibinfo {pages} {31} (\bibinfo {year} {2004})}\BibitemShut {NoStop}%
\bibitem [{\citenamefont {Schindler}\ and\ \citenamefont
  {Scherer}(2007)}]{Schindler:2006jq}%
  \BibitemOpen
  \bibfield  {author} {\bibinfo {author} {\bibfnamefont {M.~R.}\ \bibnamefont
  {Schindler}}\ and\ \bibinfo {author} {\bibfnamefont {S.}~\bibnamefont
  {Scherer}},\ }\href {\doibase 10.1140/epja/i2006-10403-3} {\bibfield
  {journal} {\bibinfo  {journal} {Eur. Phys. J.}\ }\textbf {\bibinfo {volume}
  {A32}},\ \bibinfo {pages} {429} (\bibinfo {year} {2007})}\BibitemShut
  {NoStop}%
\bibitem [{\citenamefont {Bernard}\ and\ \citenamefont
  {Meissner}(2006)}]{Bernard:2006te}%
  \BibitemOpen
  \bibfield  {author} {\bibinfo {author} {\bibfnamefont {V.}~\bibnamefont
  {Bernard}}\ and\ \bibinfo {author} {\bibfnamefont {U.-G.}\ \bibnamefont
  {Meissner}},\ }\href {\doibase 10.1016/j.physletb.2006.06.018} {\bibfield
  {journal} {\bibinfo  {journal} {Phys. Lett.}\ }\textbf {\bibinfo {volume}
  {B639}},\ \bibinfo {pages} {278} (\bibinfo {year} {2006})}\BibitemShut
  {NoStop}%
\bibitem [{\citenamefont {Procura}\ \emph {et~al.}(2007)\citenamefont
  {Procura}, \citenamefont {Musch}, \citenamefont {Hemmert},\ and\
  \citenamefont {Weise}}]{Procura:2006gq}%
  \BibitemOpen
  \bibfield  {author} {\bibinfo {author} {\bibfnamefont {M.}~\bibnamefont
  {Procura}}, \bibinfo {author} {\bibfnamefont {B.~U.}\ \bibnamefont {Musch}},
  \bibinfo {author} {\bibfnamefont {T.~R.}\ \bibnamefont {Hemmert}}, \ and\
  \bibinfo {author} {\bibfnamefont {W.}~\bibnamefont {Weise}},\ }\href
  {\doibase 10.1103/PhysRevD.75.014503} {\bibfield  {journal} {\bibinfo
  {journal} {Phys. Rev.}\ }\textbf {\bibinfo {volume} {D75}},\ \bibinfo {pages}
  {014503} (\bibinfo {year} {2007})}\BibitemShut {NoStop}%
\bibitem [{\citenamefont {Yamazaki}\ \emph {et~al.}(2009)\citenamefont
  {Yamazaki} \emph {et~al.}}]{Yamazaki:2009zq}%
  \BibitemOpen
  \bibfield  {author} {\bibinfo {author} {\bibfnamefont {T.}~\bibnamefont
  {Yamazaki}} \emph {et~al.},\ }\href {\doibase 10.1103/PhysRevD.79.114505}
  {\bibfield  {journal} {\bibinfo  {journal} {Phys. Rev.}\ }\textbf {\bibinfo
  {volume} {D79}},\ \bibinfo {pages} {114505} (\bibinfo {year}
  {2009})}\BibitemShut {NoStop}%
\bibitem [{\citenamefont {Bratt}\ \emph {et~al.}(2010)\citenamefont {Bratt}
  \emph {et~al.}}]{Bratt:2010jn}%
  \BibitemOpen
  \bibfield  {author} {\bibinfo {author} {\bibfnamefont {J.~D.}\ \bibnamefont
  {Bratt}} \emph {et~al.} (\bibinfo {collaboration} {LHPC}),\ }\href {\doibase
  10.1103/PhysRevD.82.094502} {\bibfield  {journal} {\bibinfo  {journal} {Phys.
  Rev.}\ }\textbf {\bibinfo {volume} {D82}},\ \bibinfo {pages} {094502}
  (\bibinfo {year} {2010})}\BibitemShut {NoStop}%
\bibitem [{\citenamefont {Alexandrou}\ \emph {et~al.}(2011)\citenamefont
  {Alexandrou} \emph {et~al.}}]{Alexandrou:2010hf}%
  \BibitemOpen
  \bibfield  {author} {\bibinfo {author} {\bibfnamefont {C.}~\bibnamefont
  {Alexandrou}} \emph {et~al.} (\bibinfo {collaboration} {ETM}),\ }\href
  {\doibase 10.1103/PhysRevD.83.045010} {\bibfield  {journal} {\bibinfo
  {journal} {Phys. Rev.}\ }\textbf {\bibinfo {volume} {D83}},\ \bibinfo {pages}
  {045010} (\bibinfo {year} {2011})}\BibitemShut {NoStop}%
\bibitem [{\citenamefont {Alkofer}\ and\ \citenamefont {von
  Smekal}(2001)}]{Alkofer:2000wg}%
  \BibitemOpen
  \bibfield  {author} {\bibinfo {author} {\bibfnamefont {R.}~\bibnamefont
  {Alkofer}}\ and\ \bibinfo {author} {\bibfnamefont {L.}~\bibnamefont {von
  Smekal}},\ }\href {\doibase 10.1016/S0370-1573(01)00010-2} {\bibfield
  {journal} {\bibinfo  {journal} {Phys. Rept.}\ }\textbf {\bibinfo {volume}
  {353}},\ \bibinfo {pages} {281} (\bibinfo {year} {2001})}\BibitemShut
  {NoStop}%
\bibitem [{\citenamefont {Fischer}(2006)}]{Fischer:2006ub}%
  \BibitemOpen
  \bibfield  {author} {\bibinfo {author} {\bibfnamefont {C.~S.}\ \bibnamefont
  {Fischer}},\ }\href {\doibase 10.1088/0954-3899/32/8/R02} {\bibfield
  {journal} {\bibinfo  {journal} {J. Phys.}\ }\textbf {\bibinfo {volume}
  {G32}},\ \bibinfo {pages} {R253} (\bibinfo {year} {2006})}\BibitemShut
  {NoStop}%
\bibitem [{\citenamefont {Chang}\ \emph
  {et~al.}(2011{\natexlab{a}})\citenamefont {Chang}, \citenamefont {Roberts},\
  and\ \citenamefont {Tandy}}]{Chang:2011vu}%
  \BibitemOpen
  \bibfield  {author} {\bibinfo {author} {\bibfnamefont {L.}~\bibnamefont
  {Chang}}, \bibinfo {author} {\bibfnamefont {C.~D.}\ \bibnamefont {Roberts}},
  \ and\ \bibinfo {author} {\bibfnamefont {P.~C.}\ \bibnamefont {Tandy}},\
  }\href@noop {} {\ }\Eprint {http://arxiv.org/abs/1107.4003} {1107.4003
  [nucl-th]} \BibitemShut {NoStop}%
\bibitem [{\citenamefont {Hellstern}\ \emph {et~al.}(1997)\citenamefont
  {Hellstern}, \citenamefont {Alkofer}, \citenamefont {Oettel},\ and\
  \citenamefont {Reinhardt}}]{Hellstern:1997pg}%
  \BibitemOpen
  \bibfield  {author} {\bibinfo {author} {\bibfnamefont {G.}~\bibnamefont
  {Hellstern}}, \bibinfo {author} {\bibfnamefont {R.}~\bibnamefont {Alkofer}},
  \bibinfo {author} {\bibfnamefont {M.}~\bibnamefont {Oettel}}, \ and\ \bibinfo
  {author} {\bibfnamefont {H.}~\bibnamefont {Reinhardt}},\ }\href {\doibase
  10.1016/S0375-9474(97)00514-9} {\bibfield  {journal} {\bibinfo  {journal}
  {Nucl. Phys.}\ }\textbf {\bibinfo {volume} {A627}},\ \bibinfo {pages} {679}
  (\bibinfo {year} {1997})}\BibitemShut {NoStop}%
\bibitem [{\citenamefont {Oettel}\ \emph
  {et~al.}(2000{\natexlab{a}})\citenamefont {Oettel}, \citenamefont {Alkofer},\
  and\ \citenamefont {von Smekal}}]{Oettel:2000jj}%
  \BibitemOpen
  \bibfield  {author} {\bibinfo {author} {\bibfnamefont {M.}~\bibnamefont
  {Oettel}}, \bibinfo {author} {\bibfnamefont {R.}~\bibnamefont {Alkofer}}, \
  and\ \bibinfo {author} {\bibfnamefont {L.}~\bibnamefont {von Smekal}},\
  }\href {\doibase 10.1007/s100500070078} {\bibfield  {journal} {\bibinfo
  {journal} {Eur. Phys. J.}\ }\textbf {\bibinfo {volume} {A8}},\ \bibinfo
  {pages} {553} (\bibinfo {year} {2000}{\natexlab{a}})}\BibitemShut {NoStop}%
\bibitem [{\citenamefont {Oettel}\ \emph
  {et~al.}(2000{\natexlab{b}})\citenamefont {Oettel}, \citenamefont
  {Pichowsky},\ and\ \citenamefont {von Smekal}}]{Oettel:1999gc}%
  \BibitemOpen
  \bibfield  {author} {\bibinfo {author} {\bibfnamefont {M.}~\bibnamefont
  {Oettel}}, \bibinfo {author} {\bibfnamefont {M.}~\bibnamefont {Pichowsky}}, \
  and\ \bibinfo {author} {\bibfnamefont {L.}~\bibnamefont {von Smekal}},\
  }\href {\doibase 10.1007/s100530050034} {\bibfield  {journal} {\bibinfo
  {journal} {Eur. Phys. J.}\ }\textbf {\bibinfo {volume} {A8}},\ \bibinfo
  {pages} {251} (\bibinfo {year} {2000}{\natexlab{b}})}\BibitemShut {NoStop}%
\bibitem [{\citenamefont {Ishii}(2001)}]{Ishii:2000zy}%
  \BibitemOpen
  \bibfield  {author} {\bibinfo {author} {\bibfnamefont {N.}~\bibnamefont
  {Ishii}},\ }\href {\doibase 10.1016/S0375-9474(00)00693-X} {\bibfield
  {journal} {\bibinfo  {journal} {Nucl. Phys.}\ }\textbf {\bibinfo {volume}
  {A689}},\ \bibinfo {pages} {793} (\bibinfo {year} {2001})}\BibitemShut
  {NoStop}%
\bibitem [{\citenamefont {Eichmann}\ \emph
  {et~al.}(2010{\natexlab{a}})\citenamefont {Eichmann}, \citenamefont
  {Alkofer}, \citenamefont {Krassnigg},\ and\ \citenamefont
  {Nicmorus}}]{Eichmann:2009qa}%
  \BibitemOpen
  \bibfield  {author} {\bibinfo {author} {\bibfnamefont {G.}~\bibnamefont
  {Eichmann}}, \bibinfo {author} {\bibfnamefont {R.}~\bibnamefont {Alkofer}},
  \bibinfo {author} {\bibfnamefont {A.}~\bibnamefont {Krassnigg}}, \ and\
  \bibinfo {author} {\bibfnamefont {D.}~\bibnamefont {Nicmorus}},\ }\href
  {\doibase 10.1103/PhysRevLett.104.201601} {\bibfield  {journal} {\bibinfo
  {journal} {Phys. Rev. Lett.}\ }\textbf {\bibinfo {volume} {104}},\ \bibinfo
  {pages} {201601} (\bibinfo {year} {2010}{\natexlab{a}})}\BibitemShut
  {NoStop}%
\bibitem [{\citenamefont {Sanchis-Alepuz}\ \emph {et~al.}(2011)\citenamefont
  {Sanchis-Alepuz}, \citenamefont {Eichmann}, \citenamefont {Villalba-Chavez},\
  and\ \citenamefont {Alkofer}}]{SanchisAlepuz:2011jn}%
  \BibitemOpen
  \bibfield  {author} {\bibinfo {author} {\bibfnamefont {H.}~\bibnamefont
  {Sanchis-Alepuz}}, \bibinfo {author} {\bibfnamefont {G.}~\bibnamefont
  {Eichmann}}, \bibinfo {author} {\bibfnamefont {S.}~\bibnamefont
  {Villalba-Chavez}}, \ and\ \bibinfo {author} {\bibfnamefont {R.}~\bibnamefont
  {Alkofer}},\ }\href {\doibase 10.1103/PhysRevD.84.096003} {\bibfield
  {journal} {\bibinfo  {journal} {Phys. Rev.}\ }\textbf {\bibinfo {volume}
  {D84}},\ \bibinfo {pages} {096003} (\bibinfo {year} {2011})}\BibitemShut
  {NoStop}%
\bibitem [{\citenamefont {Eichmann}(2011)}]{Eichmann:2011vu}%
  \BibitemOpen
  \bibfield  {author} {\bibinfo {author} {\bibfnamefont {G.}~\bibnamefont
  {Eichmann}},\ }\href {\doibase 10.1103/PhysRevD.84.014014} {\bibfield
  {journal} {\bibinfo  {journal} {Phys. Rev.}\ }\textbf {\bibinfo {volume}
  {D84}},\ \bibinfo {pages} {014014} (\bibinfo {year} {2011})}\BibitemShut
  {NoStop}%
\bibitem [{\citenamefont {Eichmann}\ \emph
  {et~al.}(2010{\natexlab{b}})\citenamefont {Eichmann}, \citenamefont
  {Alkofer}, \citenamefont {Krassnigg},\ and\ \citenamefont
  {Nicmorus}}]{Eichmann:2009en}%
  \BibitemOpen
  \bibfield  {author} {\bibinfo {author} {\bibfnamefont {G.}~\bibnamefont
  {Eichmann}}, \bibinfo {author} {\bibfnamefont {R.}~\bibnamefont {Alkofer}},
  \bibinfo {author} {\bibfnamefont {A.}~\bibnamefont {Krassnigg}}, \ and\
  \bibinfo {author} {\bibfnamefont {D.}~\bibnamefont {Nicmorus}},\ }\href
  {\doibase 10.1051/epjconf/20100303028} {\bibfield  {journal} {\bibinfo
  {journal} {EPJ Web Conf.}\ }\textbf {\bibinfo {volume} {3}},\ \bibinfo
  {pages} {03028} (\bibinfo {year} {2010}{\natexlab{b}})}\BibitemShut {NoStop}%
\bibitem [{\citenamefont {Maris}\ \emph {et~al.}(1998)\citenamefont {Maris},
  \citenamefont {Roberts},\ and\ \citenamefont {Tandy}}]{Maris:1997hd}%
  \BibitemOpen
  \bibfield  {author} {\bibinfo {author} {\bibfnamefont {P.}~\bibnamefont
  {Maris}}, \bibinfo {author} {\bibfnamefont {C.~D.}\ \bibnamefont {Roberts}},
  \ and\ \bibinfo {author} {\bibfnamefont {P.~C.}\ \bibnamefont {Tandy}},\
  }\href {\doibase 10.1016/S0370-2693(97)01535-9} {\bibfield  {journal}
  {\bibinfo  {journal} {Phys. Lett.}\ }\textbf {\bibinfo {volume} {B420}},\
  \bibinfo {pages} {267} (\bibinfo {year} {1998})}\BibitemShut {NoStop}%
\bibitem [{\citenamefont {Holl}\ \emph {et~al.}(2004)\citenamefont {Holl},
  \citenamefont {Krassnigg},\ and\ \citenamefont {Roberts}}]{Holl:2004fr}%
  \BibitemOpen
  \bibfield  {author} {\bibinfo {author} {\bibfnamefont {A.}~\bibnamefont
  {Holl}}, \bibinfo {author} {\bibfnamefont {A.}~\bibnamefont {Krassnigg}}, \
  and\ \bibinfo {author} {\bibfnamefont {C.~D.}\ \bibnamefont {Roberts}},\
  }\href {\doibase 10.1103/PhysRevC.70.042203} {\bibfield  {journal} {\bibinfo
  {journal} {Phys. Rev.}\ }\textbf {\bibinfo {volume} {C70}},\ \bibinfo {pages}
  {042203} (\bibinfo {year} {2004})}\BibitemShut {NoStop}%
\bibitem [{\citenamefont {Watson}\ \emph {et~al.}(2004)\citenamefont {Watson},
  \citenamefont {Cassing},\ and\ \citenamefont {Tandy}}]{Watson:2004kd}%
  \BibitemOpen
  \bibfield  {author} {\bibinfo {author} {\bibfnamefont {P.}~\bibnamefont
  {Watson}}, \bibinfo {author} {\bibfnamefont {W.}~\bibnamefont {Cassing}}, \
  and\ \bibinfo {author} {\bibfnamefont {P.~C.}\ \bibnamefont {Tandy}},\ }\href
  {\doibase 10.1007/s00601-004-0067-x} {\bibfield  {journal} {\bibinfo
  {journal} {Few Body Syst.}\ }\textbf {\bibinfo {volume} {35}},\ \bibinfo
  {pages} {129} (\bibinfo {year} {2004})}\BibitemShut {NoStop}%
\bibitem [{\citenamefont {Alkofer}\ \emph {et~al.}(2009)\citenamefont
  {Alkofer}, \citenamefont {Fischer}, \citenamefont {Llanes-Estrada},\ and\
  \citenamefont {Schwenzer}}]{Alkofer:2008tt}%
  \BibitemOpen
  \bibfield  {author} {\bibinfo {author} {\bibfnamefont {R.}~\bibnamefont
  {Alkofer}}, \bibinfo {author} {\bibfnamefont {C.~S.}\ \bibnamefont
  {Fischer}}, \bibinfo {author} {\bibfnamefont {F.~J.}\ \bibnamefont
  {Llanes-Estrada}}, \ and\ \bibinfo {author} {\bibfnamefont {K.}~\bibnamefont
  {Schwenzer}},\ }\href {\doibase 10.1016/j.aop.2008.07.001} {\bibfield
  {journal} {\bibinfo  {journal} {Annals Phys.}\ }\textbf {\bibinfo {volume}
  {324}},\ \bibinfo {pages} {106} (\bibinfo {year} {2009})}\BibitemShut
  {NoStop}%
\bibitem [{\citenamefont {Alkofer}\ \emph {et~al.}(2008)\citenamefont
  {Alkofer}, \citenamefont {Fischer},\ and\ \citenamefont
  {Williams}}]{Alkofer:2008et}%
  \BibitemOpen
  \bibfield  {author} {\bibinfo {author} {\bibfnamefont {R.}~\bibnamefont
  {Alkofer}}, \bibinfo {author} {\bibfnamefont {C.~S.}\ \bibnamefont
  {Fischer}}, \ and\ \bibinfo {author} {\bibfnamefont {R.}~\bibnamefont
  {Williams}},\ }\href {\doibase 10.1140/epja/i2008-10646-x} {\bibfield
  {journal} {\bibinfo  {journal} {Eur. Phys. J.}\ }\textbf {\bibinfo {volume}
  {A38}},\ \bibinfo {pages} {53} (\bibinfo {year} {2008})}\BibitemShut
  {NoStop}%
\bibitem [{\citenamefont {Fischer}\ and\ \citenamefont
  {Williams}(2009)}]{Fischer:2009jm}%
  \BibitemOpen
  \bibfield  {author} {\bibinfo {author} {\bibfnamefont {C.~S.}\ \bibnamefont
  {Fischer}}\ and\ \bibinfo {author} {\bibfnamefont {R.}~\bibnamefont
  {Williams}},\ }\href {\doibase 10.1103/PhysRevLett.103.122001} {\bibfield
  {journal} {\bibinfo  {journal} {Phys. Rev. Lett.}\ }\textbf {\bibinfo
  {volume} {103}},\ \bibinfo {pages} {122001} (\bibinfo {year}
  {2009})}\BibitemShut {NoStop}%
\bibitem [{\citenamefont {Fischer}\ and\ \citenamefont
  {Williams}(2008)}]{Fischer:2008wy}%
  \BibitemOpen
  \bibfield  {author} {\bibinfo {author} {\bibfnamefont {C.~S.}\ \bibnamefont
  {Fischer}}\ and\ \bibinfo {author} {\bibfnamefont {R.}~\bibnamefont
  {Williams}},\ }\href {\doibase 10.1103/PhysRevD.78.074006} {\bibfield
  {journal} {\bibinfo  {journal} {Phys. Rev.}\ }\textbf {\bibinfo {volume}
  {D78}},\ \bibinfo {pages} {074006} (\bibinfo {year} {2008})}\BibitemShut
  {NoStop}%
\bibitem [{\citenamefont {Chang}\ and\ \citenamefont
  {Roberts}(2009)}]{Chang:2009zb}%
  \BibitemOpen
  \bibfield  {author} {\bibinfo {author} {\bibfnamefont {L.}~\bibnamefont
  {Chang}}\ and\ \bibinfo {author} {\bibfnamefont {C.~D.}\ \bibnamefont
  {Roberts}},\ }\href {\doibase 10.1103/PhysRevLett.103.081601} {\bibfield
  {journal} {\bibinfo  {journal} {Phys. Rev. Lett.}\ }\textbf {\bibinfo
  {volume} {103}},\ \bibinfo {pages} {081601} (\bibinfo {year}
  {2009})}\BibitemShut {NoStop}%
\bibitem [{\citenamefont {Chang}\ \emph
  {et~al.}(2011{\natexlab{b}})\citenamefont {Chang}, \citenamefont {Liu},\ and\
  \citenamefont {Roberts}}]{Chang:2010hb}%
  \BibitemOpen
  \bibfield  {author} {\bibinfo {author} {\bibfnamefont {L.}~\bibnamefont
  {Chang}}, \bibinfo {author} {\bibfnamefont {Y.-X.}\ \bibnamefont {Liu}}, \
  and\ \bibinfo {author} {\bibfnamefont {C.~D.}\ \bibnamefont {Roberts}},\
  }\href {\doibase 10.1103/PhysRevLett.106.072001} {\bibfield  {journal}
  {\bibinfo  {journal} {Phys. Rev. Lett.}\ }\textbf {\bibinfo {volume} {106}},\
  \bibinfo {pages} {072001} (\bibinfo {year} {2011}{\natexlab{b}})}\BibitemShut
  {NoStop}%
\bibitem [{\citenamefont {Eichmann}\ \emph {et~al.}(2008)\citenamefont
  {Eichmann}, \citenamefont {Alkofer}, \citenamefont {Cloet}, \citenamefont
  {Krassnigg},\ and\ \citenamefont {Roberts}}]{Eichmann:2008ae}%
  \BibitemOpen
  \bibfield  {author} {\bibinfo {author} {\bibfnamefont {G.}~\bibnamefont
  {Eichmann}}, \bibinfo {author} {\bibfnamefont {R.}~\bibnamefont {Alkofer}},
  \bibinfo {author} {\bibfnamefont {I.~C.}\ \bibnamefont {Cloet}}, \bibinfo
  {author} {\bibfnamefont {A.}~\bibnamefont {Krassnigg}}, \ and\ \bibinfo
  {author} {\bibfnamefont {C.~D.}\ \bibnamefont {Roberts}},\ }\href {\doibase
  10.1103/PhysRevC.77.042202} {\bibfield  {journal} {\bibinfo  {journal} {Phys.
  Rev.}\ }\textbf {\bibinfo {volume} {C77}},\ \bibinfo {pages} {042202}
  (\bibinfo {year} {2008})}\BibitemShut {NoStop}%
\bibitem [{\citenamefont {Maris}\ and\ \citenamefont
  {Tandy}(1999)}]{Maris:1999nt}%
  \BibitemOpen
  \bibfield  {author} {\bibinfo {author} {\bibfnamefont {P.}~\bibnamefont
  {Maris}}\ and\ \bibinfo {author} {\bibfnamefont {P.~C.}\ \bibnamefont
  {Tandy}},\ }\href {\doibase 10.1103/PhysRevC.60.055214} {\bibfield  {journal}
  {\bibinfo  {journal} {Phys. Rev.}\ }\textbf {\bibinfo {volume} {C60}},\
  \bibinfo {pages} {055214} (\bibinfo {year} {1999})}\BibitemShut {NoStop}%
\bibitem [{\citenamefont {Qin}\ \emph {et~al.}(2011)\citenamefont {Qin},
  \citenamefont {Chang}, \citenamefont {Liu}, \citenamefont {Roberts},\ and\
  \citenamefont {Wilson}}]{Qin:2011dd}%
  \BibitemOpen
  \bibfield  {author} {\bibinfo {author} {\bibfnamefont {S.-x.}\ \bibnamefont
  {Qin}}, \bibinfo {author} {\bibfnamefont {L.}~\bibnamefont {Chang}}, \bibinfo
  {author} {\bibfnamefont {Y.-x.}\ \bibnamefont {Liu}}, \bibinfo {author}
  {\bibfnamefont {C.~D.}\ \bibnamefont {Roberts}}, \ and\ \bibinfo {author}
  {\bibfnamefont {D.~J.}\ \bibnamefont {Wilson}},\ }\href@noop {} {\bibfield
  {journal} {\bibinfo  {journal} {Phys. Rev.}\ }\textbf {\bibinfo {volume}
  {C84}},\ \bibinfo {pages} {042202} (\bibinfo {year} {2011})}\BibitemShut
  {NoStop}%
\bibitem [{\citenamefont {Maris}\ and\ \citenamefont
  {Tandy}(2006)}]{Maris:2005tt}%
  \BibitemOpen
  \bibfield  {author} {\bibinfo {author} {\bibfnamefont {P.}~\bibnamefont
  {Maris}}\ and\ \bibinfo {author} {\bibfnamefont {P.~C.}\ \bibnamefont
  {Tandy}},\ }\href {\doibase 10.1016/j.nuclphysbps.2006.08.012} {\bibfield
  {journal} {\bibinfo  {journal} {Nucl. Phys. Proc. Suppl.}\ }\textbf {\bibinfo
  {volume} {161}},\ \bibinfo {pages} {136} (\bibinfo {year}
  {2006})}\BibitemShut {NoStop}%
\bibitem [{\citenamefont {Maris}(2007)}]{Maris:2006ea}%
  \BibitemOpen
  \bibfield  {author} {\bibinfo {author} {\bibfnamefont {P.}~\bibnamefont
  {Maris}},\ }\href {\doibase 10.1063/1.2714348} {\bibfield  {journal}
  {\bibinfo  {journal} {AIP Conf. Proc.}\ }\textbf {\bibinfo {volume} {892}},\
  \bibinfo {pages} {65} (\bibinfo {year} {2007})}\BibitemShut {NoStop}%
\bibitem [{\citenamefont {Krassnigg}(2009)}]{Krassnigg:2009zh}%
  \BibitemOpen
  \bibfield  {author} {\bibinfo {author} {\bibfnamefont {A.}~\bibnamefont
  {Krassnigg}},\ }\href {\doibase 10.1103/PhysRevD.80.114010} {\bibfield
  {journal} {\bibinfo  {journal} {Phys. Rev.}\ }\textbf {\bibinfo {volume}
  {D80}},\ \bibinfo {pages} {114010} (\bibinfo {year} {2009})}\BibitemShut
  {NoStop}%
\bibitem [{\citenamefont {Nicmorus}\ \emph {et~al.}(2011)\citenamefont
  {Nicmorus}, \citenamefont {Eichmann}, \citenamefont {Krassnigg},\ and\
  \citenamefont {Alkofer}}]{Nicmorus:2010mc}%
  \BibitemOpen
  \bibfield  {author} {\bibinfo {author} {\bibfnamefont {D.}~\bibnamefont
  {Nicmorus}}, \bibinfo {author} {\bibfnamefont {G.}~\bibnamefont {Eichmann}},
  \bibinfo {author} {\bibfnamefont {A.}~\bibnamefont {Krassnigg}}, \ and\
  \bibinfo {author} {\bibfnamefont {R.}~\bibnamefont {Alkofer}},\ }\href
  {\doibase 10.1007/s00601-010-0194-5} {\bibfield  {journal} {\bibinfo
  {journal} {Few Body Syst.}\ }\textbf {\bibinfo {volume} {49}},\ \bibinfo
  {pages} {255} (\bibinfo {year} {2011})}\BibitemShut {NoStop}%
\bibitem [{\citenamefont {Haberzettl}(1997)}]{Haberzettl:1997jg}%
  \BibitemOpen
  \bibfield  {author} {\bibinfo {author} {\bibfnamefont {H.}~\bibnamefont
  {Haberzettl}},\ }\href {\doibase 10.1103/PhysRevC.56.2041} {\bibfield
  {journal} {\bibinfo  {journal} {Phys. Rev.}\ }\textbf {\bibinfo {volume}
  {C56}},\ \bibinfo {pages} {2041} (\bibinfo {year} {1997})}\BibitemShut
  {NoStop}%
\bibitem [{\citenamefont {Kvinikhidze}\ and\ \citenamefont
  {Blankleider}(1999{\natexlab{a}})}]{Kvinikhidze:1998xn}%
  \BibitemOpen
  \bibfield  {author} {\bibinfo {author} {\bibfnamefont {A.~N.}\ \bibnamefont
  {Kvinikhidze}}\ and\ \bibinfo {author} {\bibfnamefont {B.}~\bibnamefont
  {Blankleider}},\ }\href {\doibase 10.1103/PhysRevC.60.044003} {\bibfield
  {journal} {\bibinfo  {journal} {Phys. Rev.}\ }\textbf {\bibinfo {volume}
  {C60}},\ \bibinfo {pages} {044003} (\bibinfo {year}
  {1999}{\natexlab{a}})}\BibitemShut {NoStop}%
\bibitem [{\citenamefont {Kvinikhidze}\ and\ \citenamefont
  {Blankleider}(1999{\natexlab{b}})}]{Kvinikhidze:1999xp}%
  \BibitemOpen
  \bibfield  {author} {\bibinfo {author} {\bibfnamefont {A.~N.}\ \bibnamefont
  {Kvinikhidze}}\ and\ \bibinfo {author} {\bibfnamefont {B.}~\bibnamefont
  {Blankleider}},\ }\href {\doibase 10.1103/PhysRevC.60.044004} {\bibfield
  {journal} {\bibinfo  {journal} {Phys. Rev.}\ }\textbf {\bibinfo {volume}
  {C60}},\ \bibinfo {pages} {044004} (\bibinfo {year}
  {1999}{\natexlab{b}})}\BibitemShut {NoStop}%
\bibitem [{\citenamefont {Oettel}(2000)}]{Oettel:2000ig}%
  \BibitemOpen
  \bibfield  {author} {\bibinfo {author} {\bibfnamefont {M.}~\bibnamefont
  {Oettel}},\ }\href@noop {} {\ }\bibinfo {note} {PhD thesis, University of
  T\"ubingen},\ \Eprint {http://arxiv.org/abs/nucl-th/0012067}
  {nucl-th/0012067} \BibitemShut {NoStop}%
\bibitem [{\citenamefont {Mader}\ \emph {et~al.}(2011)\citenamefont {Mader},
  \citenamefont {Eichmann}, \citenamefont {Blank},\ and\ \citenamefont
  {Krassnigg}}]{Mader:2011zf}%
  \BibitemOpen
  \bibfield  {author} {\bibinfo {author} {\bibfnamefont {V.}~\bibnamefont
  {Mader}}, \bibinfo {author} {\bibfnamefont {G.}~\bibnamefont {Eichmann}},
  \bibinfo {author} {\bibfnamefont {M.}~\bibnamefont {Blank}}, \ and\ \bibinfo
  {author} {\bibfnamefont {A.}~\bibnamefont {Krassnigg}},\ }\href {\doibase
  10.1103/PhysRevD.84.034012} {\bibfield  {journal} {\bibinfo  {journal} {Phys.
  Rev.}\ }\textbf {\bibinfo {volume} {D84}},\ \bibinfo {pages} {034012}
  (\bibinfo {year} {2011})}\BibitemShut {NoStop}%
\bibitem [{\citenamefont {Kvinikhidze}\ and\ \citenamefont
  {Blankleider}(2007)}]{Kvinikhidze:2004dy}%
  \BibitemOpen
  \bibfield  {author} {\bibinfo {author} {\bibfnamefont {A.~N.}\ \bibnamefont
  {Kvinikhidze}}\ and\ \bibinfo {author} {\bibfnamefont {B.}~\bibnamefont
  {Blankleider}},\ }\href {\doibase 10.1016/j.nuclphysa.2006.11.161} {\bibfield
   {journal} {\bibinfo  {journal} {Nucl. Phys.}\ }\textbf {\bibinfo {volume}
  {A784}},\ \bibinfo {pages} {259} (\bibinfo {year} {2007})}\BibitemShut
  {NoStop}%
\bibitem [{\citenamefont {Eichmann}\ and\ \citenamefont
  {Fischer}(2011)}]{Eichmann:2011ec}%
  \BibitemOpen
  \bibfield  {author} {\bibinfo {author} {\bibfnamefont {G.}~\bibnamefont
  {Eichmann}}\ and\ \bibinfo {author} {\bibfnamefont {C.~S.}\ \bibnamefont
  {Fischer}},\ }\href@noop {} {\ }\Eprint {http://arxiv.org/abs/1111.0197}
  {1111.0197 [hep-ph]} \BibitemShut {NoStop}%
\bibitem [{\citenamefont {Maris}\ and\ \citenamefont
  {Tandy}(2000)}]{Maris:1999bh}%
  \BibitemOpen
  \bibfield  {author} {\bibinfo {author} {\bibfnamefont {P.}~\bibnamefont
  {Maris}}\ and\ \bibinfo {author} {\bibfnamefont {P.~C.}\ \bibnamefont
  {Tandy}},\ }\href {\doibase 10.1103/PhysRevC.61.045202} {\bibfield  {journal}
  {\bibinfo  {journal} {Phys. Rev.}\ }\textbf {\bibinfo {volume} {C61}},\
  \bibinfo {pages} {045202} (\bibinfo {year} {2000})}\BibitemShut {NoStop}%
\bibitem [{\citenamefont {Holl}\ \emph {et~al.}(2005)\citenamefont {Holl},
  \citenamefont {Krassnigg}, \citenamefont {Maris}, \citenamefont {Roberts},\
  and\ \citenamefont {Wright}}]{Holl:2005vu}%
  \BibitemOpen
  \bibfield  {author} {\bibinfo {author} {\bibfnamefont {A.}~\bibnamefont
  {Holl}}, \bibinfo {author} {\bibfnamefont {A.}~\bibnamefont {Krassnigg}},
  \bibinfo {author} {\bibfnamefont {P.}~\bibnamefont {Maris}}, \bibinfo
  {author} {\bibfnamefont {C.~D.}\ \bibnamefont {Roberts}}, \ and\ \bibinfo
  {author} {\bibfnamefont {S.~V.}\ \bibnamefont {Wright}},\ }\href {\doibase
  10.1103/PhysRevC.71.065204} {\bibfield  {journal} {\bibinfo  {journal} {Phys.
  Rev.}\ }\textbf {\bibinfo {volume} {C71}},\ \bibinfo {pages} {065204}
  (\bibinfo {year} {2005})}\BibitemShut {NoStop}%
\bibitem [{\citenamefont {Blank}(2011)}]{Blank:2011qk}%
  \BibitemOpen
  \bibfield  {author} {\bibinfo {author} {\bibfnamefont {M.}~\bibnamefont
  {Blank}},\ }\href@noop {} {\ }\bibinfo {note} {PhD thesis, University of
  Graz},\ \Eprint {http://arxiv.org/abs/1106.4843} {1106.4843 [hep-ph]}
  \BibitemShut {NoStop}%
\bibitem [{\citenamefont {Ball}\ and\ \citenamefont
  {Chiu}(1980)}]{Ball:1980ay}%
  \BibitemOpen
  \bibfield  {author} {\bibinfo {author} {\bibfnamefont {J.~S.}\ \bibnamefont
  {Ball}}\ and\ \bibinfo {author} {\bibfnamefont {T.-W.}\ \bibnamefont
  {Chiu}},\ }\href {\doibase 10.1103/PhysRevD.22.2542} {\bibfield  {journal}
  {\bibinfo  {journal} {Phys. Rev.}\ }\textbf {\bibinfo {volume} {D22}},\
  \bibinfo {pages} {2542} (\bibinfo {year} {1980})}\BibitemShut {NoStop}%
\bibitem [{\citenamefont {Munczek}(1995)}]{Munczek:1994zz}%
  \BibitemOpen
  \bibfield  {author} {\bibinfo {author} {\bibfnamefont {H.~J.}\ \bibnamefont
  {Munczek}},\ }\href {\doibase 10.1103/PhysRevD.52.4736} {\bibfield  {journal}
  {\bibinfo  {journal} {Phys. Rev.}\ }\textbf {\bibinfo {volume} {D52}},\
  \bibinfo {pages} {4736} (\bibinfo {year} {1995})}\BibitemShut {NoStop}%
\bibitem [{\citenamefont {Bender}\ \emph {et~al.}(1996)\citenamefont {Bender},
  \citenamefont {Roberts},\ and\ \citenamefont {Von~Smekal}}]{Bender:1996bb}%
  \BibitemOpen
  \bibfield  {author} {\bibinfo {author} {\bibfnamefont {A.}~\bibnamefont
  {Bender}}, \bibinfo {author} {\bibfnamefont {C.~D.}\ \bibnamefont {Roberts}},
  \ and\ \bibinfo {author} {\bibfnamefont {L.}~\bibnamefont {Von~Smekal}},\
  }\href {\doibase 10.1016/0370-2693(96)00372-3} {\bibfield  {journal}
  {\bibinfo  {journal} {Phys. Lett.}\ }\textbf {\bibinfo {volume} {B380}},\
  \bibinfo {pages} {7} (\bibinfo {year} {1996})}\BibitemShut {NoStop}%
\bibitem [{\citenamefont {Scherer}(2003)}]{Scherer:2002tk}%
  \BibitemOpen
  \bibfield  {author} {\bibinfo {author} {\bibfnamefont {S.}~\bibnamefont
  {Scherer}},\ }\href@noop {} {\bibfield  {journal} {\bibinfo  {journal} {Adv.
  Nucl. Phys.}\ }\textbf {\bibinfo {volume} {27}},\ \bibinfo {pages} {277}
  (\bibinfo {year} {2003})}\BibitemShut {NoStop}%
\bibitem [{\citenamefont {Khan}\ \emph {et~al.}(2006)\citenamefont {Khan} \emph
  {et~al.}}]{Khan:2006de}%
  \BibitemOpen
  \bibfield  {author} {\bibinfo {author} {\bibfnamefont {A.~A.}\ \bibnamefont
  {Khan}} \emph {et~al.},\ }\href {\doibase 10.1103/PhysRevD.74.094508}
  {\bibfield  {journal} {\bibinfo  {journal} {Phys. Rev.}\ }\textbf {\bibinfo
  {volume} {D74}},\ \bibinfo {pages} {094508} (\bibinfo {year}
  {2006})}\BibitemShut {NoStop}%
\bibitem [{\citenamefont {Nakamura}\ \emph {et~al.}(2010)\citenamefont
  {Nakamura} \emph {et~al.}}]{Nakamura:2010zzi}%
  \BibitemOpen
  \bibfield  {author} {\bibinfo {author} {\bibfnamefont {K.}~\bibnamefont
  {Nakamura}} \emph {et~al.} (\bibinfo {collaboration} {Particle Data Group}),\
  }\href {\doibase 10.1088/0954-3899/37/7A/075021} {\bibfield  {journal}
  {\bibinfo  {journal} {J. Phys.}\ }\textbf {\bibinfo {volume} {G37}},\
  \bibinfo {pages} {075021} (\bibinfo {year} {2010})}\BibitemShut {NoStop}%
\bibitem [{\citenamefont {Jenkins}\ and\ \citenamefont
  {Manohar}(1991)}]{Jenkins:1991es}%
  \BibitemOpen
  \bibfield  {author} {\bibinfo {author} {\bibfnamefont {E.~E.}\ \bibnamefont
  {Jenkins}}\ and\ \bibinfo {author} {\bibfnamefont {A.~V.}\ \bibnamefont
  {Manohar}},\ }\href {\doibase 10.1016/0370-2693(91)90840-M} {\bibfield
  {journal} {\bibinfo  {journal} {Phys. Lett.}\ }\textbf {\bibinfo {volume}
  {B259}},\ \bibinfo {pages} {353} (\bibinfo {year} {1991})}\BibitemShut
  {NoStop}%
\bibitem [{\citenamefont {Detmold}\ \emph {et~al.}(2002)\citenamefont
  {Detmold}, \citenamefont {Melnitchouk},\ and\ \citenamefont
  {Thomas}}]{Detmold:2002nf}%
  \BibitemOpen
  \bibfield  {author} {\bibinfo {author} {\bibfnamefont {W.}~\bibnamefont
  {Detmold}}, \bibinfo {author} {\bibfnamefont {W.}~\bibnamefont
  {Melnitchouk}}, \ and\ \bibinfo {author} {\bibfnamefont {A.~W.}\ \bibnamefont
  {Thomas}},\ }\href {\doibase 10.1103/PhysRevD.66.054501} {\bibfield
  {journal} {\bibinfo  {journal} {Phys. Rev.}\ }\textbf {\bibinfo {volume}
  {D66}},\ \bibinfo {pages} {054501} (\bibinfo {year} {2002})}\BibitemShut
  {NoStop}%
\bibitem [{\citenamefont {Hemmert}\ \emph {et~al.}(2003)\citenamefont
  {Hemmert}, \citenamefont {Procura},\ and\ \citenamefont
  {Weise}}]{Hemmert:2003cb}%
  \BibitemOpen
  \bibfield  {author} {\bibinfo {author} {\bibfnamefont {T.~R.}\ \bibnamefont
  {Hemmert}}, \bibinfo {author} {\bibfnamefont {M.}~\bibnamefont {Procura}}, \
  and\ \bibinfo {author} {\bibfnamefont {W.}~\bibnamefont {Weise}},\ }\href
  {\doibase 10.1103/PhysRevD.68.075009} {\bibfield  {journal} {\bibinfo
  {journal} {Phys. Rev.}\ }\textbf {\bibinfo {volume} {D68}},\ \bibinfo {pages}
  {075009} (\bibinfo {year} {2003})}\BibitemShut {NoStop}%
\bibitem [{\citenamefont {Ohta}(2011)}]{Ohta:2011nv}%
  \BibitemOpen
  \bibfield  {author} {\bibinfo {author} {\bibfnamefont {S.}~\bibnamefont
  {Ohta}} (\bibinfo {collaboration} {RBC}),\ }\href {\doibase
  10.1063/1.3587599} {\bibfield  {journal} {\bibinfo  {journal} {AIP Conf.
  Proc.}\ }\textbf {\bibinfo {volume} {1354}},\ \bibinfo {pages} {148}
  (\bibinfo {year} {2011})}\BibitemShut {NoStop}%
\bibitem [{\citenamefont {Schroder}\ \emph {et~al.}(2001)\citenamefont
  {Schroder} \emph {et~al.}}]{Schroder:2001rc}%
  \BibitemOpen
  \bibfield  {author} {\bibinfo {author} {\bibfnamefont {H.~C.}\ \bibnamefont
  {Schroder}} \emph {et~al.},\ }\href {\doibase 10.1007/s100520100754}
  {\bibfield  {journal} {\bibinfo  {journal} {Eur. Phys. J.}\ }\textbf
  {\bibinfo {volume} {C21}},\ \bibinfo {pages} {473} (\bibinfo {year}
  {2001})}\BibitemShut {NoStop}%
\bibitem [{\citenamefont {Alexandrou}\ \emph {et~al.}(2007)\citenamefont
  {Alexandrou}, \citenamefont {Koutsou}, \citenamefont {Leontiou},
  \citenamefont {Negele},\ and\ \citenamefont {Tsapalis}}]{Alexandrou:2007xj}%
  \BibitemOpen
  \bibfield  {author} {\bibinfo {author} {\bibfnamefont {C.}~\bibnamefont
  {Alexandrou}}, \bibinfo {author} {\bibfnamefont {G.}~\bibnamefont {Koutsou}},
  \bibinfo {author} {\bibfnamefont {T.}~\bibnamefont {Leontiou}}, \bibinfo
  {author} {\bibfnamefont {J.~W.}\ \bibnamefont {Negele}}, \ and\ \bibinfo
  {author} {\bibfnamefont {A.}~\bibnamefont {Tsapalis}},\ }\href {\doibase
  10.1103/PhysRevD.76.094511} {\bibfield  {journal} {\bibinfo  {journal} {Phys.
  Rev.}\ }\textbf {\bibinfo {volume} {D76}},\ \bibinfo {pages} {094511}
  (\bibinfo {year} {2007})}\BibitemShut {NoStop}%
\bibitem [{\citenamefont {Lyubushkin}\ \emph {et~al.}(2009)\citenamefont
  {Lyubushkin} \emph {et~al.}}]{Lyubushkin:2008pe}%
  \BibitemOpen
  \bibfield  {author} {\bibinfo {author} {\bibfnamefont {V.}~\bibnamefont
  {Lyubushkin}} \emph {et~al.} (\bibinfo {collaboration} {NOMAD}),\ }\href
  {\doibase 10.1140/epjc/s10052-009-1113-0} {\bibfield  {journal} {\bibinfo
  {journal} {Eur. Phys. J.}\ }\textbf {\bibinfo {volume} {C63}},\ \bibinfo
  {pages} {355} (\bibinfo {year} {2009})}\BibitemShut {NoStop}%
\bibitem [{\citenamefont {Gran}\ \emph {et~al.}(2006)\citenamefont {Gran} \emph
  {et~al.}}]{Gran:2006jn}%
  \BibitemOpen
  \bibfield  {author} {\bibinfo {author} {\bibfnamefont {R.}~\bibnamefont
  {Gran}} \emph {et~al.} (\bibinfo {collaboration} {K2K}),\ }\href {\doibase
  10.1103/PhysRevD.74.052002} {\bibfield  {journal} {\bibinfo  {journal} {Phys.
  Rev.}\ }\textbf {\bibinfo {volume} {D74}},\ \bibinfo {pages} {052002}
  (\bibinfo {year} {2006})}\BibitemShut {NoStop}%
\bibitem [{\citenamefont {Aguilar-Arevalo}\ \emph {et~al.}(2010)\citenamefont
  {Aguilar-Arevalo} \emph {et~al.}}]{AguilarArevalo:2010zc}%
  \BibitemOpen
  \bibfield  {author} {\bibinfo {author} {\bibfnamefont {A.~A.}\ \bibnamefont
  {Aguilar-Arevalo}} \emph {et~al.} (\bibinfo {collaboration} {MiniBooNE}),\
  }\href {\doibase 10.1103/PhysRevD.81.092005} {\bibfield  {journal} {\bibinfo
  {journal} {Phys. Rev.}\ }\textbf {\bibinfo {volume} {D81}},\ \bibinfo {pages}
  {092005} (\bibinfo {year} {2010})}\BibitemShut {NoStop}%
\bibitem [{\citenamefont {Nieves}\ \emph {et~al.}(2011)\citenamefont {Nieves},
  \citenamefont {Simo},\ and\ \citenamefont {Vacas}}]{Nieves:2011yp}%
  \BibitemOpen
  \bibfield  {author} {\bibinfo {author} {\bibfnamefont {J.}~\bibnamefont
  {Nieves}}, \bibinfo {author} {\bibfnamefont {I.~R.}\ \bibnamefont {Simo}}, \
  and\ \bibinfo {author} {\bibfnamefont {M.~J.~V.}\ \bibnamefont {Vacas}},\
  }\href@noop {} {\ }\Eprint {http://arxiv.org/abs/1106.5374} {1106.5374
  [hep-ph]} \BibitemShut {NoStop}%
\bibitem [{\citenamefont {Bhattacharya}\ \emph {et~al.}(2011)\citenamefont
  {Bhattacharya}, \citenamefont {Hill},\ and\ \citenamefont
  {Paz}}]{Bhattacharya:2011ah}%
  \BibitemOpen
  \bibfield  {author} {\bibinfo {author} {\bibfnamefont {B.}~\bibnamefont
  {Bhattacharya}}, \bibinfo {author} {\bibfnamefont {R.~J.}\ \bibnamefont
  {Hill}}, \ and\ \bibinfo {author} {\bibfnamefont {G.}~\bibnamefont {Paz}},\
  }\href@noop {} {\ }\Eprint {http://arxiv.org/abs/1108.0423} {1108.0423
  [hep-ph]} \BibitemShut {NoStop}%
\bibitem [{\citenamefont {Blank}\ and\ \citenamefont
  {Krassnigg}(2011)}]{Blank:2011ha}%
  \BibitemOpen
  \bibfield  {author} {\bibinfo {author} {\bibfnamefont {M.}~\bibnamefont
  {Blank}}\ and\ \bibinfo {author} {\bibfnamefont {A.}~\bibnamefont
  {Krassnigg}},\ }\href@noop {} {\ }\Eprint {http://arxiv.org/abs/1109.6509}
  {1109.6509 [hep-ph]} \BibitemShut {NoStop}%
\bibitem [{\citenamefont {Eichmann}(2009)}]{Eichmann:2009zx}%
  \BibitemOpen
  \bibfield  {author} {\bibinfo {author} {\bibfnamefont {G.}~\bibnamefont
  {Eichmann}},\ }\href@noop {} {\ }\bibinfo {note} {PhD thesis, University of
  Graz},\ \Eprint {http://arxiv.org/abs/0909.0703} {0909.0703 [hep-ph]}
  \BibitemShut {NoStop}%
\bibitem [{\citenamefont {Nicmorus}\ \emph {et~al.}(2009)\citenamefont
  {Nicmorus}, \citenamefont {Eichmann}, \citenamefont {Krassnigg},\ and\
  \citenamefont {Alkofer}}]{Nicmorus:2008vb}%
  \BibitemOpen
  \bibfield  {author} {\bibinfo {author} {\bibfnamefont {D.}~\bibnamefont
  {Nicmorus}}, \bibinfo {author} {\bibfnamefont {G.}~\bibnamefont {Eichmann}},
  \bibinfo {author} {\bibfnamefont {A.}~\bibnamefont {Krassnigg}}, \ and\
  \bibinfo {author} {\bibfnamefont {R.}~\bibnamefont {Alkofer}},\ }\href
  {\doibase 10.1103/PhysRevD.80.054028} {\bibfield  {journal} {\bibinfo
  {journal} {Phys. Rev.}\ }\textbf {\bibinfo {volume} {D80}},\ \bibinfo {pages}
  {054028} (\bibinfo {year} {2009})}\BibitemShut {NoStop}%
\bibitem [{\citenamefont {Nicmorus}\ \emph {et~al.}(2010)\citenamefont
  {Nicmorus}, \citenamefont {Eichmann},\ and\ \citenamefont
  {Alkofer}}]{Nicmorus:2010sd}%
  \BibitemOpen
  \bibfield  {author} {\bibinfo {author} {\bibfnamefont {D.}~\bibnamefont
  {Nicmorus}}, \bibinfo {author} {\bibfnamefont {G.}~\bibnamefont {Eichmann}},
  \ and\ \bibinfo {author} {\bibfnamefont {R.}~\bibnamefont {Alkofer}},\ }\href
  {\doibase 10.1103/PhysRevD.82.114017} {\bibfield  {journal} {\bibinfo
  {journal} {Phys. Rev.}\ }\textbf {\bibinfo {volume} {D82}},\ \bibinfo {pages}
  {114017} (\bibinfo {year} {2010})}\BibitemShut {NoStop}%
\bibitem [{\citenamefont {Krassnigg}\ and\ \citenamefont
  {Maris}(2005)}]{Krassnigg:2004if}%
  \BibitemOpen
  \bibfield  {author} {\bibinfo {author} {\bibfnamefont {A.}~\bibnamefont
  {Krassnigg}}\ and\ \bibinfo {author} {\bibfnamefont {P.}~\bibnamefont
  {Maris}},\ }\href {\doibase 10.1088/1742-6596/9/1/029} {\bibfield  {journal}
  {\bibinfo  {journal} {J. Phys. Conf. Ser.}\ }\textbf {\bibinfo {volume}
  {9}},\ \bibinfo {pages} {153} (\bibinfo {year} {2005})}\BibitemShut {NoStop}%
\bibitem [{\citenamefont {Krassnigg}(2008)}]{Krassnigg:2009gd}%
  \BibitemOpen
  \bibfield  {author} {\bibinfo {author} {\bibfnamefont {A.}~\bibnamefont
  {Krassnigg}},\ }\href@noop {} {\bibfield  {journal} {\bibinfo  {journal}
  {PoS}\ }\textbf {\bibinfo {volume} {CONFINEMENT8}},\ \bibinfo {pages} {075}
  (\bibinfo {year} {2008})}\BibitemShut {NoStop}%
\bibitem [{\citenamefont {Andreev}\ \emph {et~al.}(2007)\citenamefont {Andreev}
  \emph {et~al.}}]{Andreev:2007wg}%
  \BibitemOpen
  \bibfield  {author} {\bibinfo {author} {\bibfnamefont {V.~A.}\ \bibnamefont
  {Andreev}} \emph {et~al.} (\bibinfo {collaboration} {MuCap}),\ }\href
  {\doibase 10.1103/PhysRevLett.99.032002} {\bibfield  {journal} {\bibinfo
  {journal} {Phys. Rev. Lett.}\ }\textbf {\bibinfo {volume} {99}},\ \bibinfo
  {pages} {032002} (\bibinfo {year} {2007})}\BibitemShut {NoStop}%
\end{thebibliography}%

\end{document}